\documentclass{aa}  
\usepackage[varg]{txfonts}
\usepackage{graphicx}
\usepackage{amssymb,amsmath}
\usepackage{booktabs,multirow}
\usepackage{bm}
\usepackage{natbib}
\bibpunct{(}{)}{;}{a}{}{,}

\begin{document}

\title{On the use of relative field line helicity as an indicator for solar eruptivity}

\author{K. Moraitis$^1$ \and S. Patsourakos$^1$ \and A. Nindos$^1$ \and J.~K. Thalmann$^2$ \and \'{E}. Pariat$^3$}
\institute{Physics Department, University of Ioannina, Ioannina GR-45110, Greece \and University of Graz, Institute of Physics/IGAM, Graz, Austria \and Sorbonne Universit\'{e}, \'{E}cole polytechnique, Institut Polytechnique de Paris, Universit\'{e} Paris Saclay, Observatoire de Paris-PSL, CNRS, Laboratoire de Physique des Plasmas (LPP), 4 place Jussieu, 75005 Paris, France}

\date{Received ... / Accepted ...}

\abstract{Relative field line helicity (RFLH) is a recently developed quantity which can approximate the density of relative magnetic helicity.}{This paper aims to determine whether RFLH can be used as an indicator of solar eruptivity.}{Starting from magnetographic observations from the Helioseismic and Magnetic Imager instrument onboard the Solar Dynamic Observatory of a sample of seven solar active regions (ARs), which comprises over 2000 individual snapshots, we reconstruct the AR's coronal magnetic field with a widely-used non-linear force-free method. This enables us to compute RFLH using two independent gauge conditions for the vector potentials. We focus our study around the times of strong flares in the ARs, above the M class, and in regions around the polarity inversion lines (PILs) of the magnetic field, and of RFLH.}{We find that the temporal profiles of the relative helicity that is contained in the magnetic PIL follow those of the relative helicity that is computed by the accurate volume method for the whole AR. Additionally, the PIL relative helicity can be used to define a parameter similar to the well-known parameter $R$ \citep{schrijver07}, whose high values are related with increased flaring probability. This helicity-based $R$-parameter correlates well with the original one, showing in some cases even higher values, and additionally, it experiences more pronounced decreases during flares. This means that there exists at least one parameter deduced from RFLH, that has important value as a solar eruptivity indicator.}{}


\titlerunning{Relative field line helicity and eruptivity}
\authorrunning{Moraitis et al.}

\maketitle

\section{Introduction}
\label{sect:introduction}

Magnetic helicity quantifies the complexity of a magnetic field, as it measures the twist and writhe of individual magnetic field lines, and the intertwining of pairs of lines. For a magnetic field $\mathbf{B}$ in a volume $V$, it can be defined through the volume integral $H=\int_V \mathbf{A}\cdot \mathbf{B}\,{\rm d}V$, where $\mathbf{A}$ is the vector potential of $\mathbf{B}$. Magnetic helicity is a conserved quantity in ideal magnetohydrodynamics (MHD) \citep{woltjer58}, and is approximately very well conserved in non-ideal conditions when the plasma magnetic Reynolds number is high \citep{taylor74,pariat15}. It thus determines the evolution of astrophysical plasmas which obey the MHD equations. Magnetic helicity is a gauge-dependent quantity unless the boundary of the volume of interest is a flux surface, that is, magnetic flux neither enters nor leaves through it. Since this condition is not met in the case of the Sun, the appropriate quantity is then relative (magnetic) helicity \citep{BergerF84}, which is expressed with the help of a reference magnetic field.

There are situations where it is desirable to know the locations where helicity is stronger, as these are expected to be important in the flaring and eruptive processes \citep[e.g.,][]{mctaggart21}. Such spatial information cannot be provided by neither the classical, nor by the relative helicity due to the non-locality of the vector potentials and the resulting inability to define the respective densities, which, moreover, are gauge dependent as follows from the definition of helicity. A meaningful proxy for the density of magnetic helicity is field line helicity \citep{berger88}. Field line helicity (FLH) expresses the magnetic helicity of a field line per unit of magnetic flux. It is unique for each field line and it has a clear physical meaning for closed field lines, while it is gauge-dependent for open field lines \citep{yeates16}, field lines that close outside of the volume of interest, that is.

Field line helicity can be generalized so that to include the case of relative helicity following two distinct formulations, given in \citet{yeates18} and \citet{moraitis19}, which differ in the gauges of the vector potentials they use. These relative field line helicities (RFLHs), when weighted by the magnetic flux and summed over all field lines that close on the boundary of the volume, reproduce relative helicity.

The classical FLH and the RFLH have been examined in various applications on the Sun, mostly in situations which approximate the solar conditions, or in MHD simulations. As an example, \citet{russel15} employed FLH in a magnetic field set-up that approximates coronal loops, in order to examine the connectivity changes caused by reconnection. In two more examples, \citet{yeates16} and \citet{lowder17} applied FLH to the global magnetic field of the Sun to study the distribution of helicity in the solar corona. The latter work was able to determine the locations of flux ropes in the corona, by setting a threshold in the values of FLH.

Other solar-oriented applications of RFLH, where the target is the large emerging eruptive active region (AR) 11158, are those of \citet{moraitis19} and \citet{moraitis21}. In both these works, non-linear force-free (NLFF) reconstructions of the coronal magnetic field were used to obtain RFLH. In the former work, the relation of RFLH with the (also gauge-dependent) helicity flux density \citep{pariat05} was examined at a specific instant, and shown that the two quantities exhibit similar photospheric distributions. In \citet{moraitis21}, a more detailed study of RFLH was carried out over an extended time interval. It was shown that RFLH was capable of associating the large decrease in the value of helicity during an X-class flare of the AR with the magnetic structure that later erupted. Moreover, the use of RFLH enabled the authors to calculate the relative helicity contained in the central part of the AR and around the flare ribbons, and to compare it with the relative helicity of the whole AR.

Magnetic helicity is in general a potentially key quantity to understand the eruptivity of solar ARs. Studies so far have focused on different aspects of magnetic helicity, such as its global accumulation and content in ARs \citep[e.g.,][]{Nindos04,Labonte07,Tziotziou12,thalmann21}, its flux \citep[e.g.,][]{Moon02,ParkSH10,liokati22,liu23}, the distribution of its flux \citep[e.g.,][]{Vemareddy12,Dalmasse18}, or the ratio of two components in its decomposition \citep[e.g.,][]{pariat17,pariat23,Zuccarello18,Linan18,moraitis19b,thalmann19,green22}. However, direct volume helicity (resp. helicity flux) measurements only provide a single scalar value of the helicity content (resp. helicity injection) at a given time for the whole magnetic system \citep[see review of][]{valori16}. They cannot help in the finer understanding of the system's dynamics in relation with the volume distribution of helicity.

Identifying adjacent magnetic flux structures of opposite helicity sign appears as especially interesting to understand the eruptive dynamics. Such systems can lead to helicity annihilation if they interact by magnetic reconnection \citep[e.g.,][]{kusano04}. Simulations of the interaction of flux tubes with different chirality have showed that magnetic reconnection can release more energy when it occurs between systems of opposite helicity \citep{linton01}. The theoretical understanding behind this result is that helicity annihilation enables the system to reach a lower energy level than it would have with constant helicity. This motivates the search for adjacent systems in which helicity has opposite signs, as is one of the regions of interest that will be analysed in the present study.

Sequential and/or simultaneous helicity injection has frequently been observed in flaring and erupting ARs \citep[e.g.,][]{Chandra10,Romano10,ParkSH12,Vemareddy17}. Using a connectivity-based helicity flux proxy, \citet{Dalmasse13} has remarked that the very subdomain of an active region that was erupting was the one associated with neighbouring magnetix flux systems of opposite helicity flux. While helicity flux is usually more straigtforward to measure observationaly \citep[e.g.,][]{Demoulin09}, it does not provide the proper helicity sign of the different magnetic domains of the active region. The helicity variations in a flux tube are probably not directly related to its global sign. All the results on the relation between eruptivity and helicity that are based on the helicity flux distribution may be missing the key element of comparison: the proper 3D distribution of magnetic helicity which is solely provided by RFLH.

Overall, all the previous results point towards the great potential that RFLH has in highlighting the locations which are involved in solar eruptivity. Moreover, the photospheric morphology of RFLH could be used to define parameters related to increased solar activity. Such parameters could help the search for solar eruptivity indicators which has reached a considerable level of maturity, with the best-performing ones showing skill scores of around 80\% \citep{barnes16,jonas18,kontogiannis19,flarecast}.

In this work, we examine whether there can be parameters derived from RFLH that indicate upcoming solar eruptivity, by studying the behaviour of RFLH in a larger sample of observed flaring ARs than in \citet{moraitis21}. In Sect.~\ref{sect:data} we describe the observational data that we used and the considered ARs, while in Sect.~\ref{sect:method} we define all quantities of interest and the methodology for deriving them from the observational data. In Sect.~\ref{sect:results} we present the results obtained from the application of RFLH in the chosen AR sample, focusing on the times around strong flares. Finally, in Sect.~\ref{sect:discussion} we summarize and discuss the results of the paper.

\section{The AR sample}
\label{sect:data}

The AR sample for the study of RFLH was chosen so that it includes a variety of behaviours regarding the level of activity and morphology of the individual ARs. The only observational data used in this work was the \texttt{sharp$\_$cea$\_$720s} data product from the Helioseismic and Magnetic Imager \citep[HMI,][]{sche12} instrument of the Solar Dynamic Observatory \citep[SDO,][]{pes12}, which consists of vector magnetograms. Some basic observational characteristics of the ARs considered in this work are given below.

\subsection{AR 11158}
\label{sect:158}

The dataset for this AR corresponds to 115 snapshots at the typical cadence of 1~h, except around its two strongest flares when it was 12~min. The observations were from 12 February 2011, 00:00 UT to 16 February 2011, 00:00 UT. During this interval two M- and an X-class flare occured. The magnetograms were rebinned to a pixel size of 2", or 1440 km, and cropped to the final dimensions of 148$\times$92 pixels. More details can be found in \citet{thalmann19}. The X-class flare was the subject of the work in \citet{moraitis21} but we include it here as well.

\subsection{AR 11261}
\label{sect:261}

We considered 60 snapshots at 12 min cadence, from 3 August 2011, 21:00 UT to 4 August 2011, 08:48 UT, when an M9.3 flare occured \citep{sarkar19}. The original magnetograms were rebinned to a pixel size of 1" (or 720 km), and then cropped to the size of 292$\times$200 pixels, with their lower left corner at the (120,30) pixel of the rebinned image. In this and all other ARs where rebinning and cropping were applied, it was checked that the two processes approximately conserved the original values of the photoshperic flux, force, and torque balance parameters \citep{wiegelmann06}.

\subsection{AR 11520}
\label{sect:520}

We considered 61 snapshots at 12 min cadence, at 12 July 2012, from 09:00 UT to 21:00 UT, when an X1.4 flare occured. The original magnetograms were rebinned to a pixel size of 1", and then cropped to the size of 528$\times$296 pixels, with their lower left corner at the (82,152) pixel.

\subsection{AR 11618}
\label{sect:618}
 
We considered 675 snapshots at 12 min cadence (except from two gaps of 36 min and 1 day), from 18 November 2012, 20:00 UT to 25 November 2012, 11:12 UT, when four M-class flares occured \citep{liokati23}. The magnetograms were rebinned to a pixel size of 1", and then cropped to the size of 384$\times$172 pixels, with their lower left corner at the (30,1) pixel.

\subsection{AR 11890}
\label{sect:890}

We considered 892 snapshots at 12 min cadence (except from two gaps of 36 and 84 min), from 5 November 2013, 14:00 UT to 13 November 2013, 01:48 UT. Four M- and three X-class flares occured then \citep{liokati23}. The magnetograms were rebinned to a pixel size of 1", and then cropped to the size of 316$\times$232 pixels, with their lower left corner at the (78,5) pixel.

\subsection{AR 12192}
\label{sect:192}

We considered 198 snapshots at the cadence of 1 h except around large flares when it was 12 min, from 20 October 2014, 12:00 UT to 25 October 2014, 03:00 UT. Nine M- and two X-class flares occured during this interval. The magnetograms were rebinned to a pixel size of 2", and cropped to the final dimensions of 276$\times$200 pixels. More details can be found in \citet{thalmann19}.

\subsection{AR 12673}
\label{sect:673}

The dataset for this AR consists of 48 snapshots at 12 min cadence (except from a 2.5 hour gap), at 6 September 2017, from 06:00 UT to 17:48 UT, when an M- and two X-class flares occured. The magnetograms were rebinned to a pixel size of 1", and then cropped to the size of 256$\times$224 pixels, with their lower left corner at the (44,0) pixel.

\section{Methodology}
\label{sect:method}

\subsection{Magnetic field modelling}

The starting point for all calculations of RFLH and all other physical quantities of interest, is the computation of the 3D coronal magnetic field. This task is the most demanding computationally. The 3D coronal field at each instant, and for each AR, is extrapolated from the corresponding observed HMI magnetogram with the NLFF method of \citet{wieg10}, which is further described in \citet{wieg12}. In all cases, the photospheric magnetic field is pre-processed and smoothed before the NLFF method is applied \citep{wiegelmann06}. Moreover, the default parameters for weighting the solenoidal and force-free terms in the code are used here. The full details of the method can be found in \citet{thalmann19} where the NLFF fields in the case of ARs 11158 and 12192 were first used.

One extrapolation parameter that differs among the various ARs is the size of the non-physical boundary layer which surrounds the physical domain in both the horizontal, and the upward vertical directions. This parameter ranges from 12 to 24 pixels, depending on the size of the physical domain of each AR, having larger values at larger domains. In the case of AR 11261 for example, this layer was taken equal to 12 pixels, so that the final coronal grid consisted of 268$\times$176$\times$176 pixels, corresponding to 190$\times$125$\times$125~Mm. Another parameter that varies among ARs is the vertical height of the extrapolation volume. This was chosen equal to the smallest of the respective horizontal spatial dimensions in five of the ARs, and to 128 pixels in the other two where the extrapolations were already available. A summary of the physical characteristics of the produced coronal magnetic fields for all ARs is given in Table~\ref{tab1}. We finally note that all subsequent computations of this work were performed on the volumes of Table~\ref{tab1}.

\begin{table}[ht]
\caption{Characteristics of the produced coronal magnetic fields.}
\centering
\begin{tabular}{ccccccc}
\hline
AR & grid points & pixel size & snapshots \\
\hline
11158 & 148$\times$92$\times$128 & 2" & 115 \\
11261 & 268$\times$176$\times$176 & 1" & 60 \\
11520 & 480$\times$248$\times$248 & 1" & 61 \\
11618 & 352$\times$140$\times$140 & 1" & 675 \\
11890 & 284$\times$200$\times$200 & 1" & 892 \\
12192 & 276$\times$200$\times$128 & 2" & 198 \\
12673 & 232$\times$200$\times$200 & 1" & 48 \\
\hline
\end{tabular}
\label{tab1}
\end{table}

The effect of different pixel sizes between ARs is further examined in Appendix~\ref{app2}. It is shown there that if we had considered the same pixel size in all ARs, the differences in our results would be on the order of 10\%. Similar levels of changes due to binning to the volume-integrated energies and helicities have also been observed in \citet{thalmann22}.

\subsection{Relative helicity and field line helicity}

The focus of this work is on magnetic helicities. We compute relative magnetic helicity in a volume $V$ \citep{BergerF84,fa85}, from its definition
\begin{equation}
H_\mathrm{r}=\int_V (\mathbf{A}+\mathbf{A}_\mathrm{p})\cdot (\mathbf{B}-\mathbf{B}_\mathrm{p})\,{\rm d}V.
\label{helr}
\end{equation}
The magnetic field $\mathbf{B}_\mathrm{p}$ is a reference field that is chosen to be potential, while $\mathbf{A}$, $\mathbf{A}_\mathrm{p}$ are the respective vector potentials of the two magnetic fields. Relative magnetic helicity is independent of the gauges of the vector potentials as long as the following condition holds
\begin{equation}
\left. \hat{n}\cdot \mathbf{B} \right|_{\partial V}=\left. \hat{n}\cdot \mathbf{B}_\mathrm{p} \right|_{\partial V},
\label{helc}
\end{equation}
where $\hat{n}$ denotes the outward-pointing unit normal on the boundary of the volume, $\partial V$.

Magnetic helicity can be computed in a two-step process when the magnetic field is given, as described in \citet{moraitis14}. The first step is the computation of the potential field that satisfies the condition of Eq.~(\ref{helc}), which results from the numerical solution of a 3D Laplace equation. The second step involves the computation of the vector potentials from the respective magnetic fields. This task is facilitated by adopting the \citet[][DV]{devore00} gauge, as described in \citet{valori16}, namely, $A_z=A_{p,z}=0$. We just note that in all our computations, $\mathbf{A}$ is in the simple DV gauge where only integrations are involved, while $\mathbf{A}_\mathrm{p}$ is in a DV gauge that, simultaneously, satisfies the Coulomb gauge \citep[Sect.~2.2 of][]{val12}.

Relative (magnetic) field line helicity, $h_r$, acts as a physically-motivated field-line density for relative helicity. It can be defined in multiple ways depending on whether we consider the positive-polarity part of the boundary ($\partial V^+$), the negative-polarity one ($\partial V^-$), or both ($\partial V$), namely
\begin{equation}
H_\mathrm{r}=\oint_{\partial V^+} h^+_\mathrm{r}\,{\rm d}\Phi=\oint_{\partial V^-} h^-_\mathrm{r}\,{\rm d}\Phi=\oint_{\partial V} h_\mathrm{r}\,{\rm d}\Phi.
\label{flhhel}
\end{equation}
Here ${\rm d}\Phi=\left| \hat{n}\cdot\mathbf{B} \right|\,{\rm d}S$ stands for the elementary magnetic flux on the boundary, and ${\rm d}S$ for the respective elementary area. In the remaining we choose the last expression as the RFLH, which can always be written as the average of its expressions at the positive-, and negative-polarity parts of the boundary, namely
\begin{equation}
h_\mathrm{r}=\frac{1}{2}\left( h_\mathrm{r}^+ + h_\mathrm{r}^-\right).
\label{flhdef0}
\end{equation}
The factor of $1/2$ ensures that each field line is accounted for only once.

The positive-, and negative-polarity RFLHs can be calculated following two independent formulations. In the first, no assumption is made for the gauges of the vector potentials, and the resulting RFLHs retain their most general expression \citep{moraitis19}. The two RFLHs are then given by
\begin{equation}
h_\mathrm{r}^+=\int_{\alpha_+}^{\alpha_-}\,(\mathbf{A}+\mathbf{A}_\mathrm{p}) \cdot {\rm d}\bm{l} - \int_{\alpha_{+}}^{\alpha_{p-}}\,(\mathbf{A}+\mathbf{A}_\mathrm{p}) \cdot {\rm d}\bm{l}_\mathrm{p},
\label{flhdef1}
\end{equation}
and
\begin{equation}
h_\mathrm{r}^-=\int_{\alpha_+}^{\alpha_-}\,(\mathbf{A}+\mathbf{A}_\mathrm{p}) \cdot {\rm d}\bm{l} - \int_{\alpha_{p+}}^{\alpha_{-}}\,(\mathbf{A}+\mathbf{A}_\mathrm{p}) \cdot {\rm d}\bm{l}_\mathrm{p}.
\label{flhdef2}
\end{equation}
Here, $\alpha_+$, $\alpha_-$ ($\alpha_{p+}$, $\alpha_{p-}$) denote the footpoints of $\mathbf{B}$ ($\mathbf{B}_\mathrm{p}$) where flux enters into, or leaves from the volume, respectively, while ${\rm d}\bm{l}$, ${\rm d}\bm{l}_\mathrm{p}$ are the elementary lengths along the respective field lines.

In practise, in the computation of Eqs.~(\ref{flhdef1}), (\ref{flhdef2}) we employ the same DV gauge combination that is used for relative helicity, and refer to it as the DV gauge, $h_\mathrm{r}^\mathrm{DV}$, in the following. The required field line integrations along the two magnetic fields are performed with the method described in \citet{moraitis19}. As in that work, we restrict the footpoints for RFLH computation on the photospheric boundary where RFLH is most important. Previous experience has shown that the few field lines that close on the lateral or top boundaries have insignificant contribution to RFLH. The numerical code (in C/C++) for the computation of RFLH in the DV gauge can be found at \url{doi.org/10.5281/zenodo.7211784}.

In the second formulation, a commonly-used choice for the gauges of the vector potentials is made from the start \citep{yeates18}, namely that
\begin{equation}
\left. \hat{n}\times\mathbf{A}_\mathrm{p} \right|_{\partial V} = \left. \hat{n}\times\mathbf{A} \right|_{\partial V}.
\label{eq:gaugeb2}
\end{equation}
Since this gauge was used in the original definition of relative helicity \citep{BergerF84}, we refer to it as the Berger $\&$ Field gauge (BF). The respective RFLH expressions are then
\begin{equation}
h_\mathrm{r}^{\mathrm{BF},+}= \int_{\alpha_+}^{\alpha_-}\,\mathbf{A} \cdot {\rm d}\bm{l} - \int_{\alpha_{+}}^{\alpha_{p-}}\,\mathbf{A}_\mathrm{p} \cdot {\rm d}\bm{l}_\mathrm{p}
\label{flhyp1}
\end{equation}
and
\begin{equation}
h_\mathrm{r}^{\mathrm{BF},-}= \int_{\alpha_+}^{\alpha_-}\,\mathbf{A} \cdot {\rm d}\bm{l} - \int_{\alpha_{p+}}^{\alpha_{-}}\,\mathbf{A}_\mathrm{p} \cdot {\rm d}\bm{l}_\mathrm{p}.
\label{flhyp2}
\end{equation}
As in the DV case, we restrict the footpoints for RFLH computation on the photospheric boundary only. In both formulations, the RFLHs are expressed as a line integral along the field lines of $\mathbf{B}$ relative to another line integral along $\mathbf{B}_\mathrm{p}$, which justifies their name as relative.

For the computation of the RFLH in the BF gauge, $h_r^\mathrm{BF}$, the three steps described above are repeated anew, since in the respective code \citep{yeates18} all vector quantities are assumed in a different grid than the ones in the DV gauge (staggered instead of collocated). This does not affect the computed RFLH, which is given in the same locations as $h_\mathrm{r}^\mathrm{DV}$, only the intermediate computational steps. The respective python/Fortran code can be found at \url{github.com/antyeates1983/flhtools}.

\section{Results}
\label{sect:results}

\subsection{RFLH morphology}
\label{sect:res1}

\begin{figure*}[ht]
\centering
\includegraphics[width=0.32\textwidth]{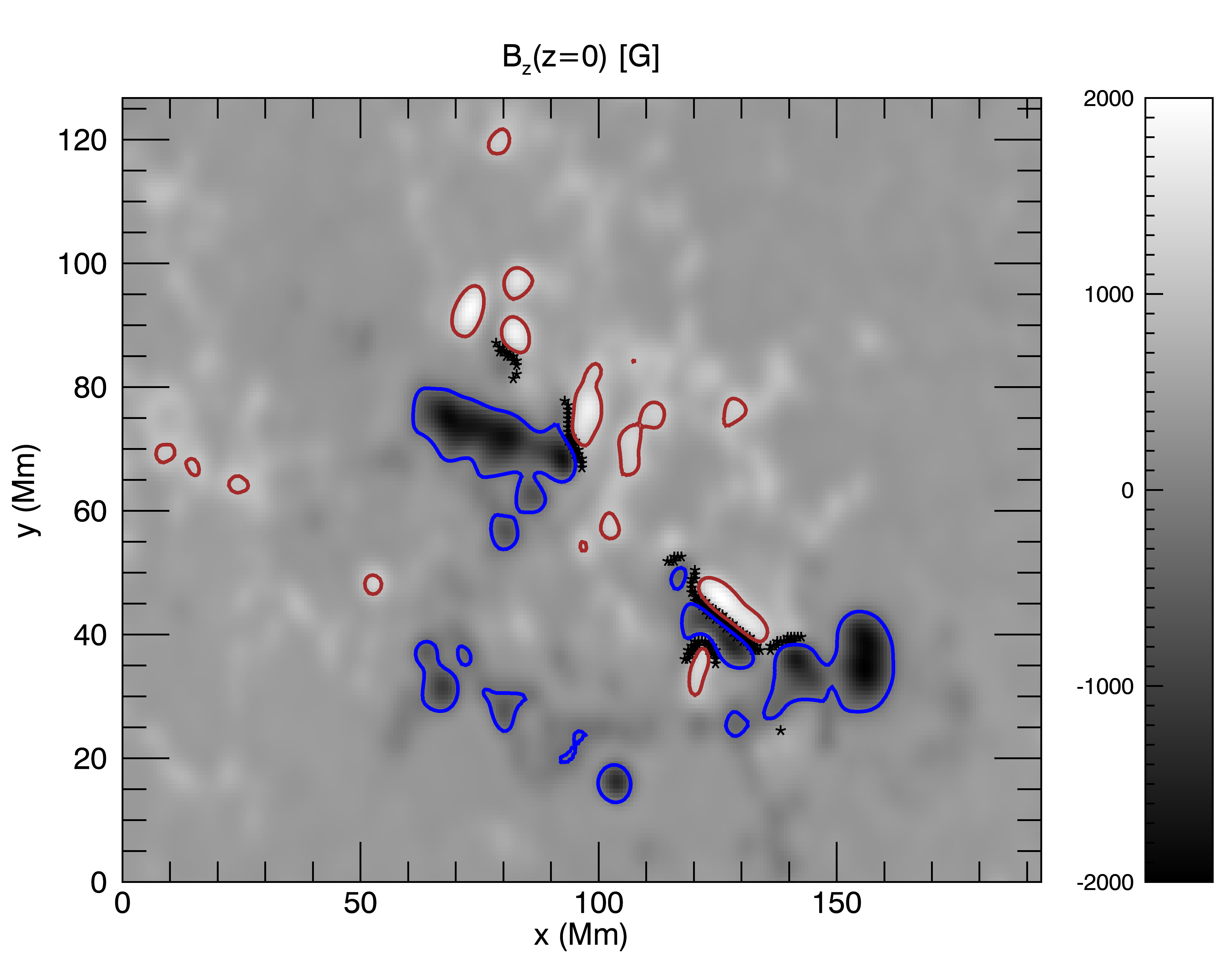}%
\includegraphics[width=0.32\textwidth]{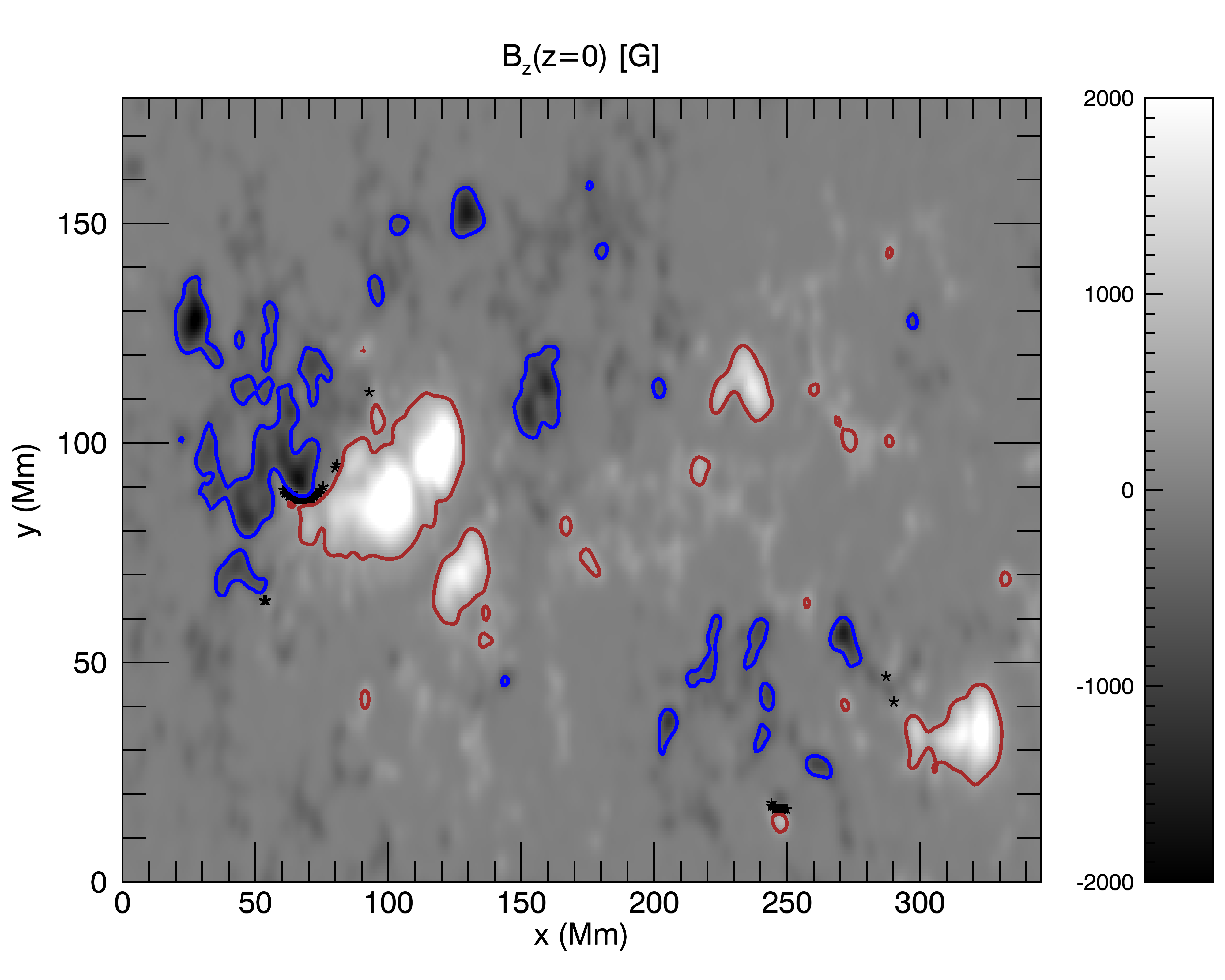}%
\includegraphics[width=0.32\textwidth]{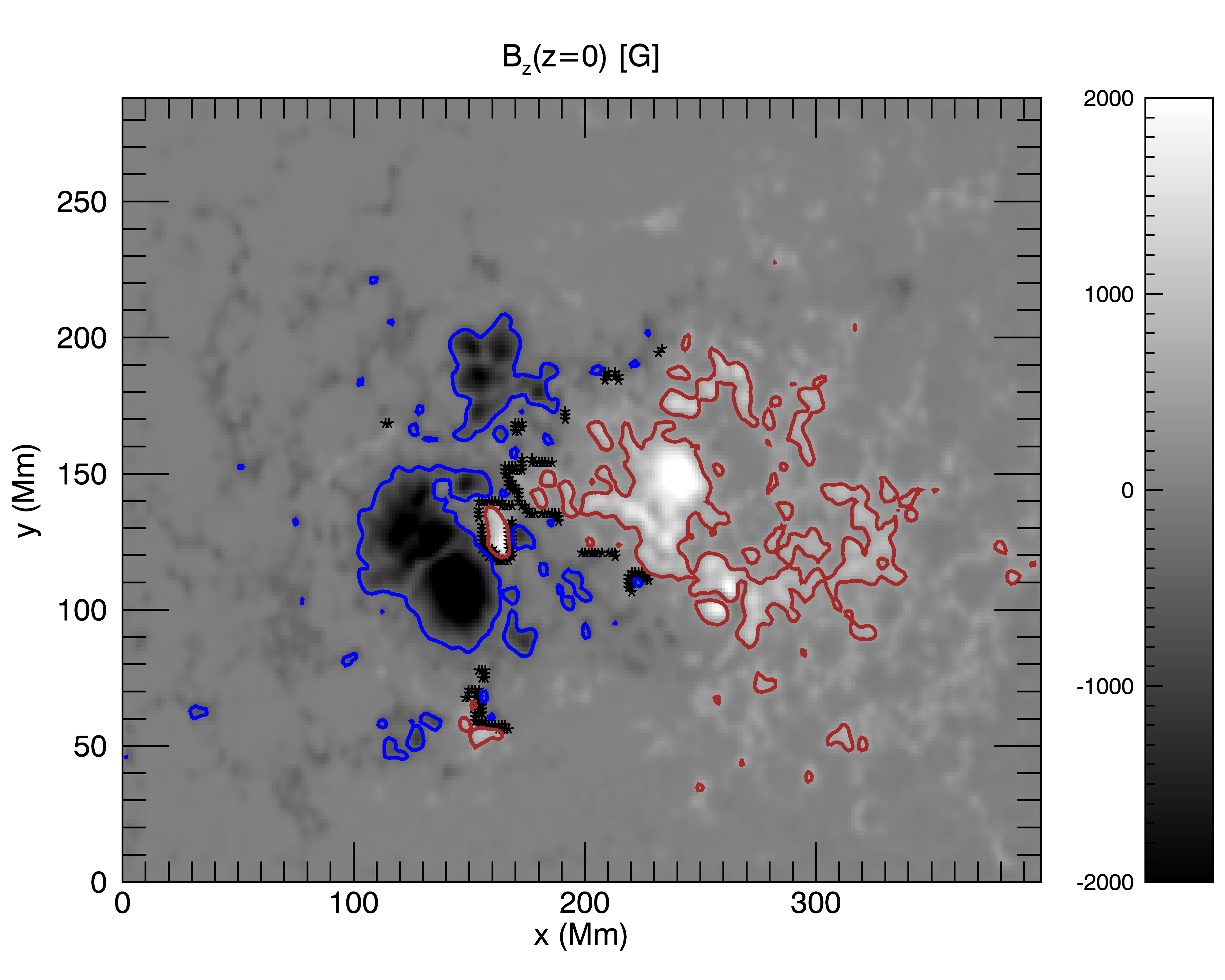}\\
\includegraphics[width=0.32\textwidth]{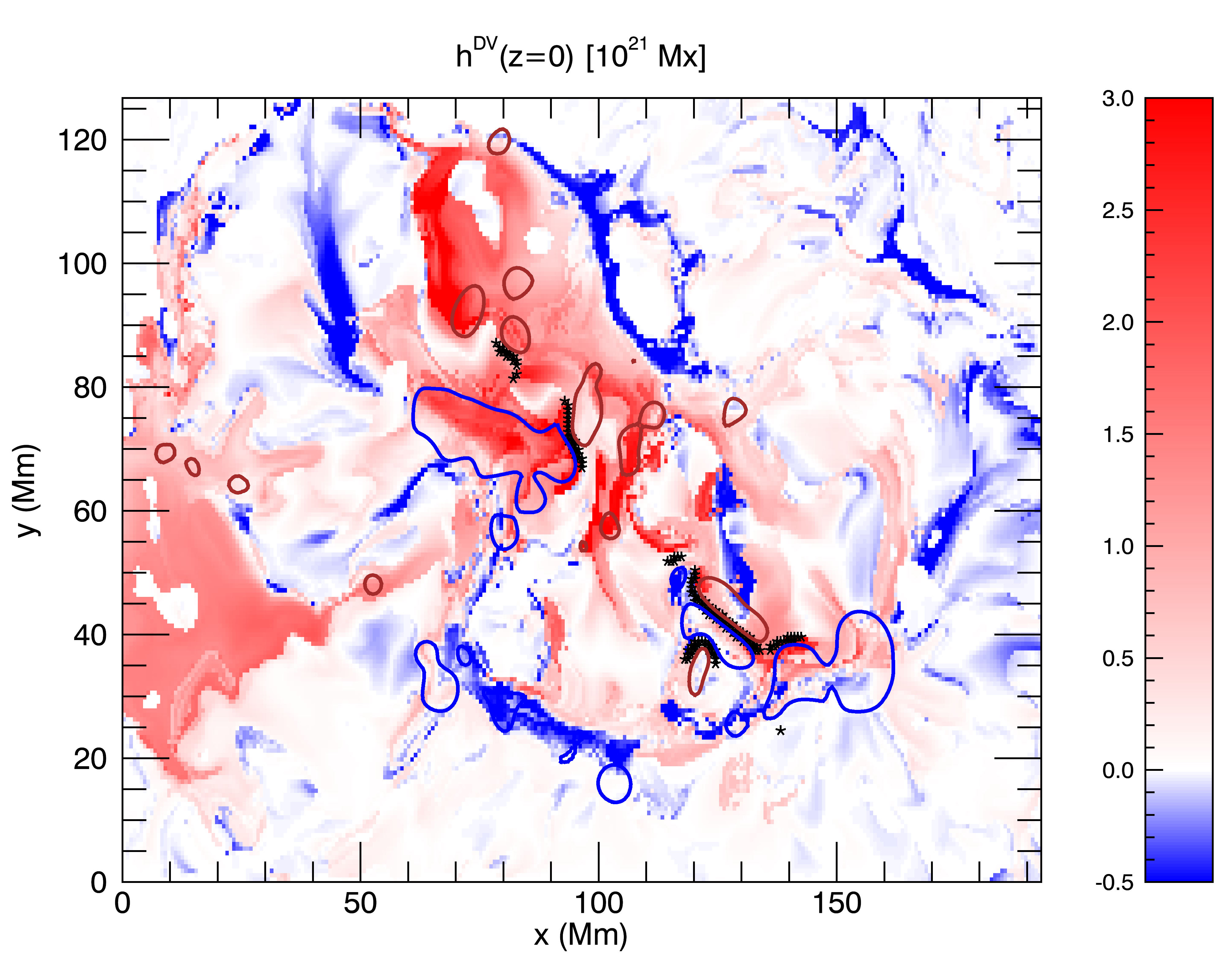}%
\includegraphics[width=0.32\textwidth]{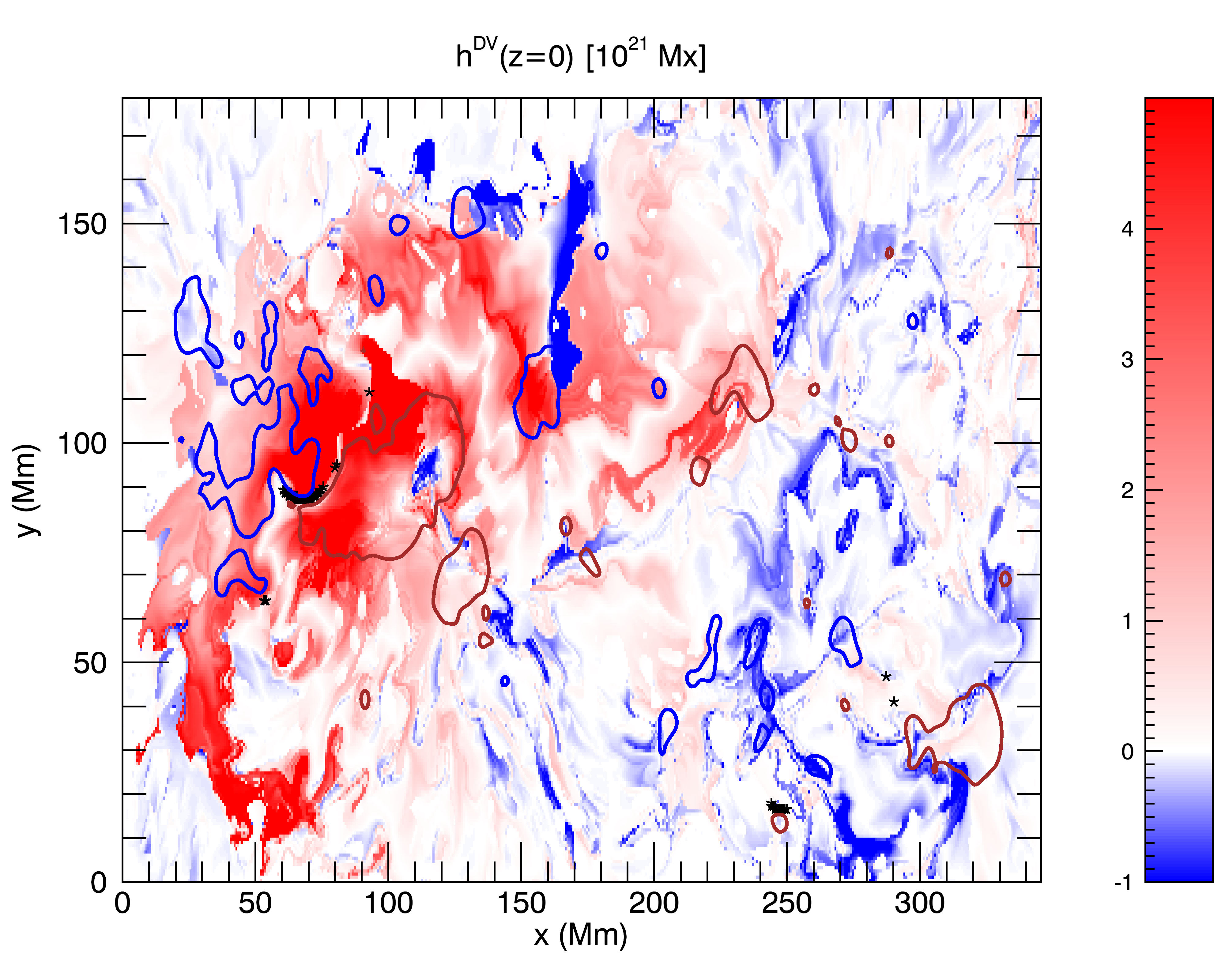}%
\includegraphics[width=0.32\textwidth]{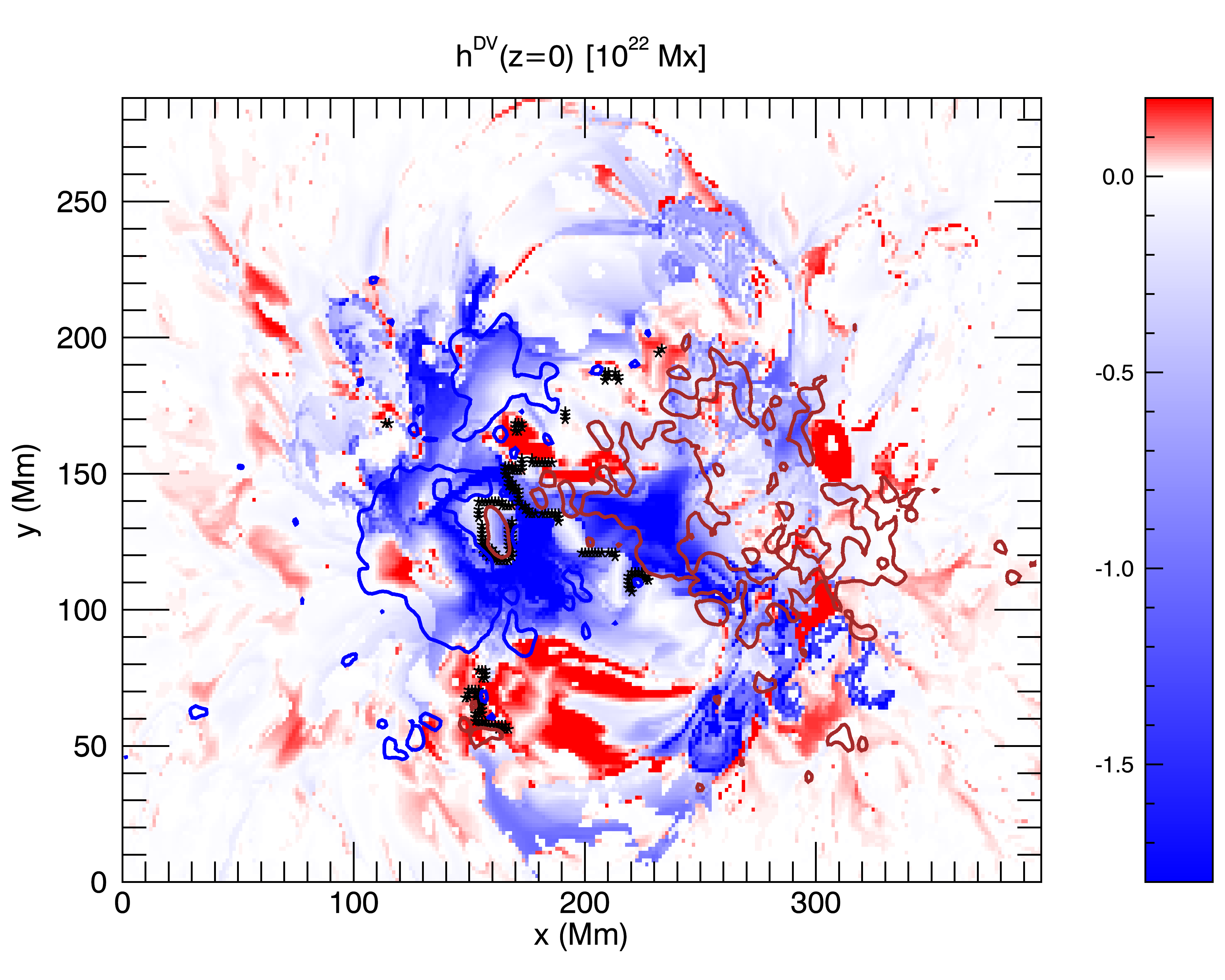}\\
\includegraphics[width=0.32\textwidth]{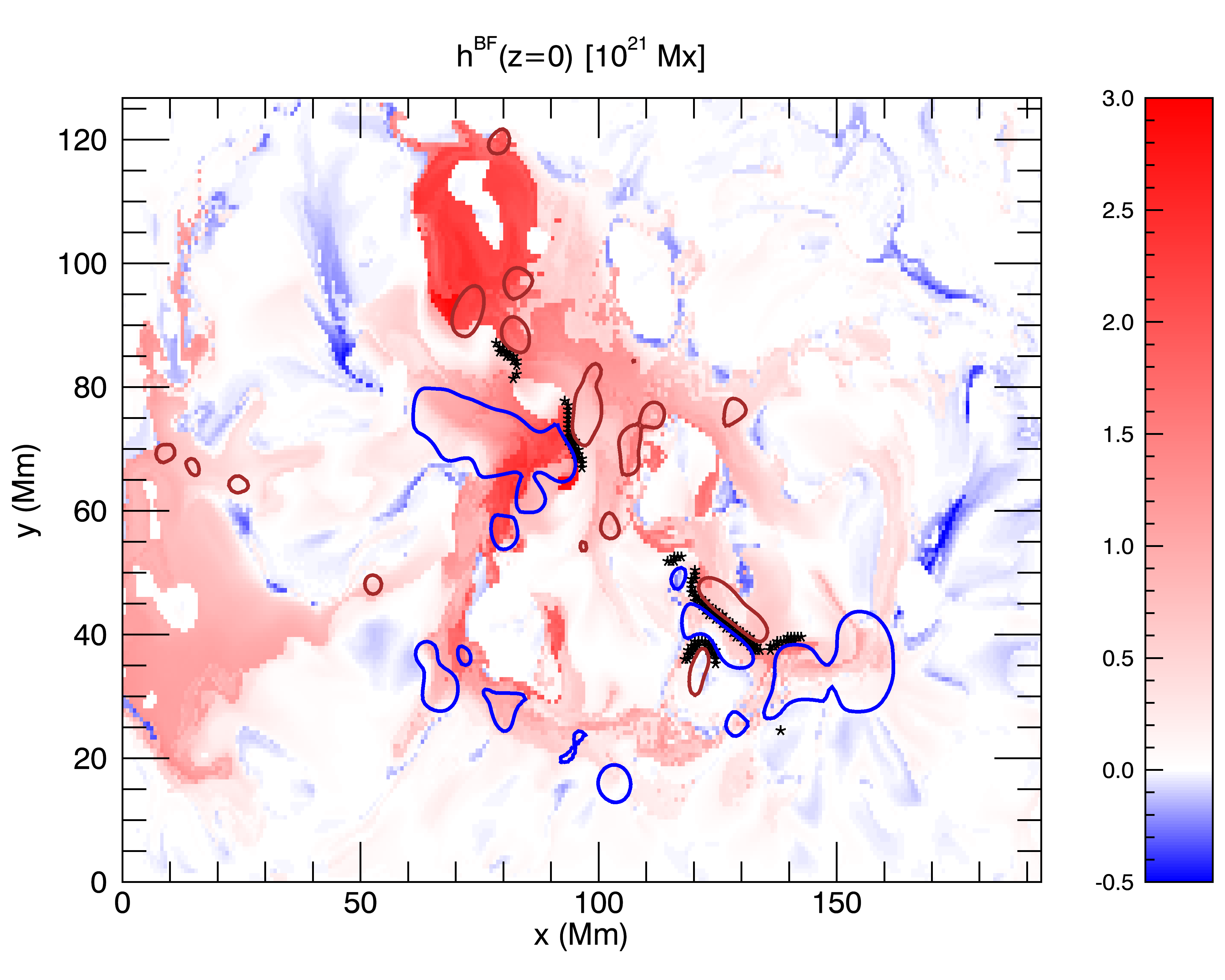}%
\includegraphics[width=0.32\textwidth]{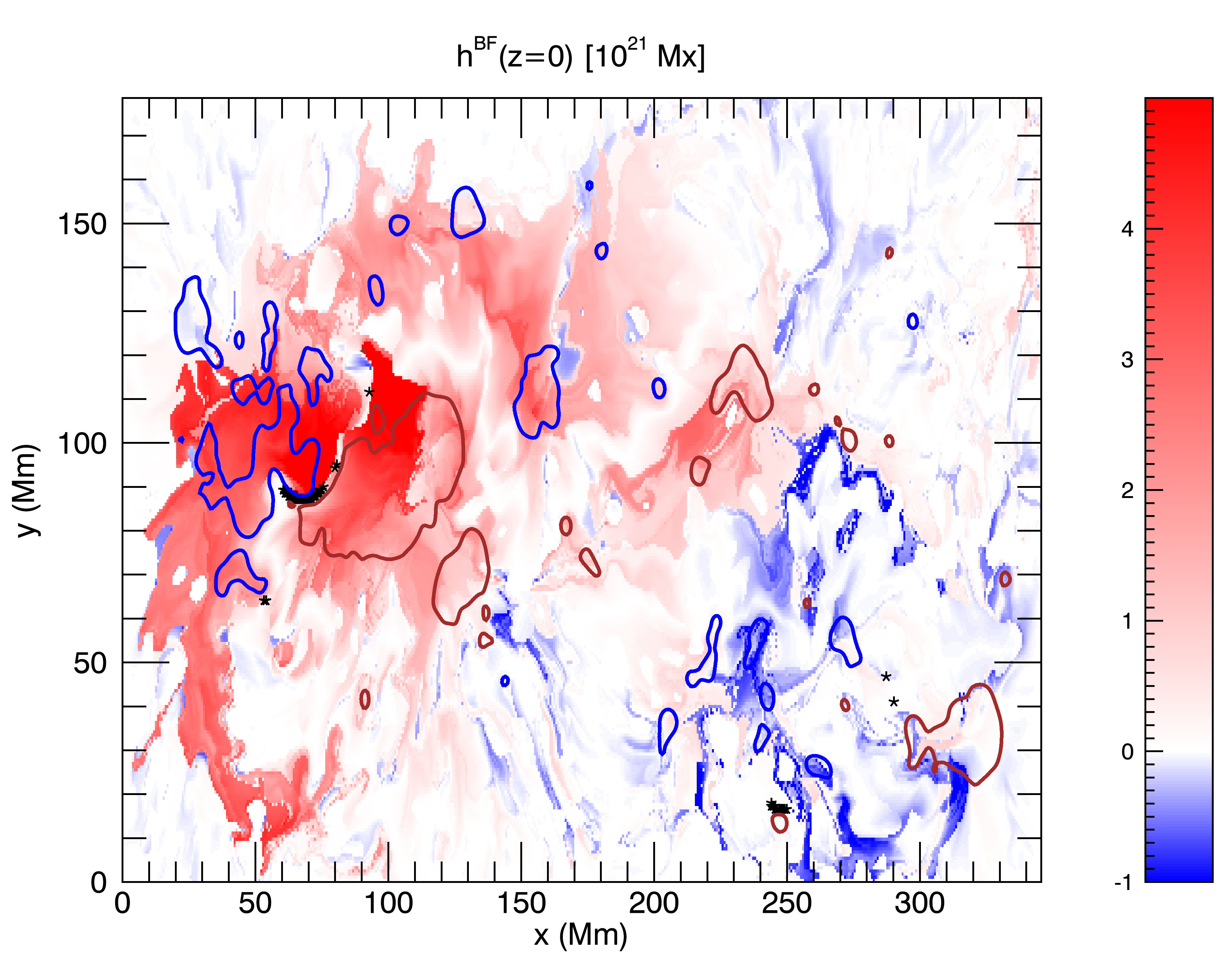}%
\includegraphics[width=0.32\textwidth]{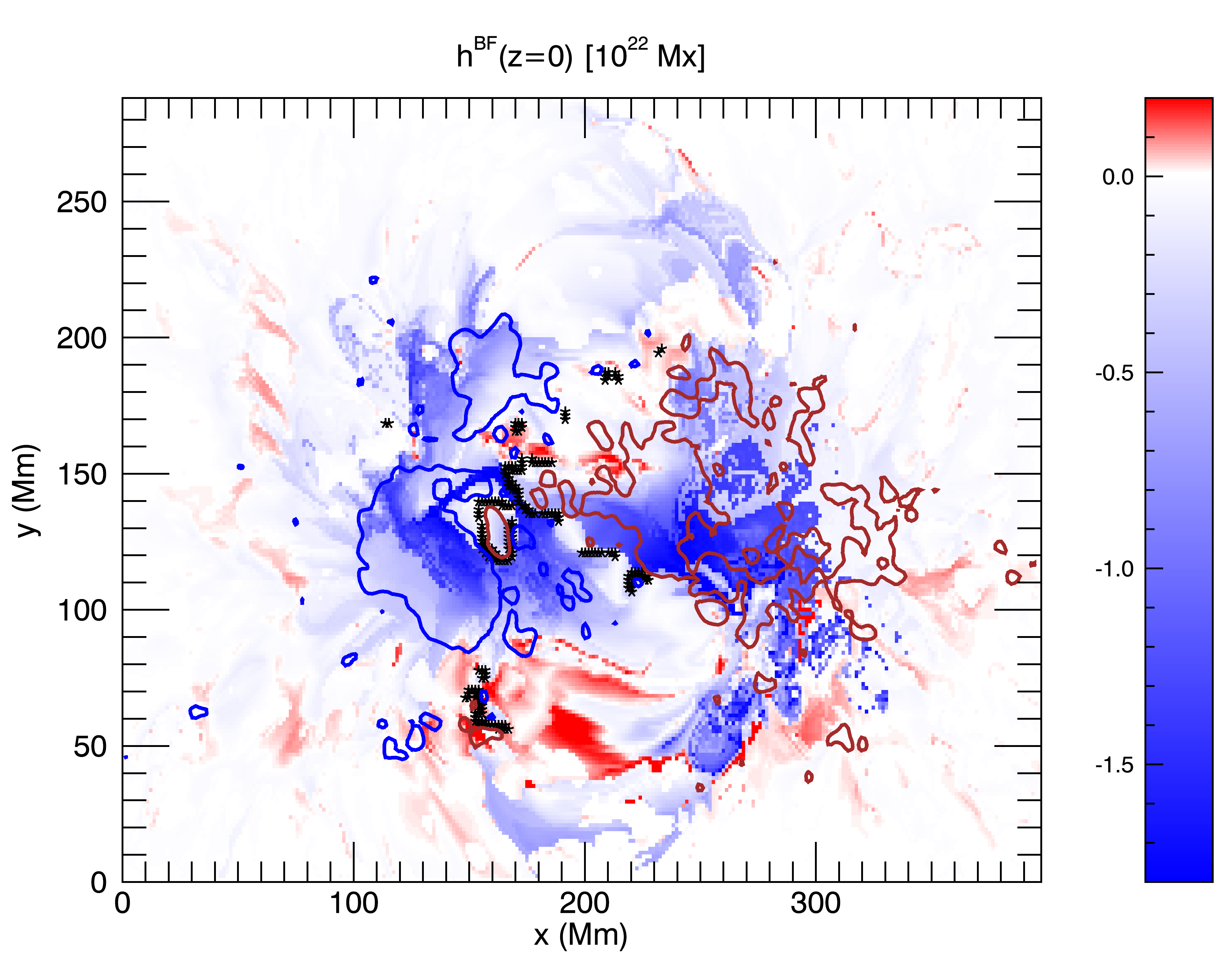}
\caption{Selected snapshots showing the photospheric distribution of the normal component of the magnetic field $B_z$ (top row), and of RFLH in the DV (middle row), and in the BF gauge (bottom row), right before the strongest flares of three ARs. For AR 11261 this was on 03:36 UT of 4 August 2011 (left column), for AR 11520 on 15:36 UT of 12 July 2012 (middle column), and for AR 12192 on 21:11 UT of 24 October 2014 (right column). Red (blue) contours correspond to values of $B_z=+500$~G ($B_z=-500$~G), while the locations of the simple magnetic polarity inversion lines are shown with black stars. From left to right, the pairs of RFLH images exhibit pixel-to-pixel linear correlation coefficients of 0.78, 0.84, and 0.75.}
\label{flhmorph}
\end{figure*}

For each of the few hundreds snapshots of all ARs, we first compute the NLFF extrapolated field, and then all quantities of interest (helicities, RFLHs), following the methodology of Sect.~\ref{sect:method}. In Fig.~\ref{flhmorph} we show three examples of the 2D distribution of RFLH on the photospheric plane ($z=0$) in three different ARs, and in both the DV and BF gauges. The three specific snapshots correspond to the times right before the start of the strongest flares of AR 11261, AR 11520, and AR 12192 (from left to right). We should remind the reader here that RFLH is a function of the field lines and so all field lines that close on the photosphere should have the same value of RFLH on both footpoints.

It is evident from Fig.~\ref{flhmorph} that the ARs we considered exhibit a large variety in behaviours, which is something we wanted to have. A first difference between the various ARs is in the sign and magnitude of RFLH, with two of the ARs in Fig.~\ref{flhmorph} having mostly positive values and the other mostly negative. The distributions of RFLH in all ARs are more diffuse compared to those of the magnetic field, with strong RFLH regions occupying much of the field of view. This is expected since magnetic helicity is highly non-local and, unlike magnetic flux, it is influenced by the large scale distribution of magnetic flux, as was already seen in models \citep{moraitis19}, and also in observed ARs \citep{moraitis21}. Similar to the latter work, RFLH can have important values outside regions of strong magnetic flux concentrations in some cases, something that could be attributed to the larger length of field lines there. More frequently however, RFLH tends to exhibit high values around the polarity inversion lines (PILs) of the magnetic field.

We also note that, in most cases, the overall morphology of RFLH is not very sensitive to the gauge used in its computation. This is especially true at the regions of highest RFLH values, a result also found in \citet{moraitis21}. The good agreement of RFLH in the two gauges is depicted in the high values of the pixel-to-pixel linear (Pearson's) correlation coefficients between them, which are on the order of 0.8, with similar values for the respective Spearman's rank correlation coefficients (SCC). The most obvious difference between the two gauges is in the magnitude of RFLH, which in the BF gauge is smaller by up to a factor of two in some cases. The RFLH distributions obtained with the two gauges exhibit similar levels of smoothness, as the Coulomb gauge is used in the computation of the vector potential $\mathbf{A}_\mathrm{p}$, in both the DV and the BF gauges. Additionally, the morphology of RFLH around the PILs is quite similar in the two gauges. The overall similarity of RFLH in the two gauges stems from the fact that RFLH is a field line integrated quantity, and thus it depicts the structure of the field line mapping, as also seen in \citet{yeates18}. All these remarks hold for the other snapshots and ARs, as well.

The RFLH data for all ARs and all snapshots, in both examined gauges, can be obtained at \url{https://zenodo.org/records/7607307}. Despite the wealth of all this information we focus in this work on specific spatio-temporal regions of the RFLH images, and leave their detailed examination for future work.

\subsection{Flare sample}
\label{sect:flares}

We start by restricting temporally our study at the times around strong flares. Three criteria are used to identify the flares. First, the flares have to be above the M1.0 class. Second, they must have the highest HMI cadence of 12 min so that the $\pm$~1h time profiles of the examined quantities contain enough points. Finally, the quality of the NLFF field at these times should be high enough. More specifically, the value of the energy ratio $E_\mathrm{div}/E$, which is a measure of the non-solenoidality of a magnetic field and is defined in Appendix~\ref{app1} along with other measures of solenoidality and force-freeness, should be $\lesssim0.05$, as proposed in \citet{thalmann19a}. After applying these criteria, the resulting flare sample consists of 22 flares from seven ARs. We note that other choices for the threshold were also examined, including the more relaxed value of $E_\mathrm{div}/E\lesssim 0.08$ which has been suggested earlier \citep{valori16}, and the results are reported in Appendix~\ref{app3}.

The properties of all flares are summarized in Table~\ref{tab2}. The times and positions of the ARs are taken mostly from the flare ribbons catalogue of \citet{kazachenko17}, but also from the `Solar Activity Archive' (\url{helio.mssl.ucl.ac.uk/helio-vo/solar_activity/arstats-archive/}). The characterization of ARs as eruptive or not is deduced from various sources, including the latter archive, and also \citet{liokati23}, \citet{liu23}, and \citet{sarkar19}.

The spatial distribution of the considered ARs on the solar disk are two in the northern hemisphere and five in the southern one. Moreover, the flares are equally split into eruptive and confined, although the majority of the latter (nine) come from a single AR, namely AR 12192. Five of the flares are of the X-, and 17 of the M-class. All X-class flares in our sample, but the one of AR 12673 (number 22), are also considered in \citet{liu23}.

\begin{table*}[ht]
\caption{Characteristics of the flare sample used in the analysis of RFLH.}
\centering
\begin{tabular}{cccccccc}
\hline
flare number & AR & flare class & eruptive (Y/N) & peak time & position & ($E_\mathrm{div}/E)\times 10^2$ \\
\hline
01 & 11158 & M6.6 & Y  & 2011-02-13/17:38UT & S20E04 & 0.23$\pm$0.03 \\
02 & 11158 & X2.2 & Y  & 2011-02-15/01:56UT & S20W10 & 0.50$\pm$0.03 \\
03 & 11261 & M9.3 & Y  & 2011-08-04/03:45UT & N19W36 & 4.73$\pm$0.07 \\
04 & 11520 & X1.4 & Y  & 2012-07-12/16:49UT & S15W01 & 3.80$\pm$0.19 \\
05 & 11618 & M1.7 & Y  & 2012-11-20/12:38UT & N06E20 & 1.96$\pm$0.13 \\
06 & 11618 & M1.6 & Y  & 2012-11-20/19:28UT & N07E15 & 0.86$\pm$0.08 \\
07 & 11618 & M1.4 & Y  & 2012-11-21/06:48UT & N06E10 & 0.53$\pm$0.07 \\
08 & 11618 & M3.5 & Y  & 2012-11-21/15:28UT & N08E14 & 0.53$\pm$0.05 \\
09 & 11890 & M1.0 & N  & 2013-11-05/18:13UT & S12E47 & 4.70$\pm$0.13 \\
10 & 11890 & M2.3 & Y  & 2013-11-07/03:40UT & S14E28 & 3.97$\pm$0.08 \\
11 & 11890 & M2.4 & Y  & 2013-11-07/14:25UT & S13E23 & 3.60$\pm$0.14 \\
12 & 12192 & M4.5 & N  & 2014-10-20/16:37UT & S14E37 & 1.98$\pm$0.08 \\
13 & 12192 & M1.4 & N  & 2014-10-20/19:02UT & S13E43 & 2.11$\pm$0.06 \\
14 & 12192 & M1.7 & N  & 2014-10-20/20:03UT & S13E43 & 1.88$\pm$0.05 \\
15 & 12192 & M1.2 & N  & 2014-10-20/22:55UT & S14E36 & 2.22$\pm$0.03 \\
16 & 12192 & M8.7 & N  & 2014-10-22/01:59UT & S14E19 & 2.07$\pm$0.03 \\
17 & 12192 & M2.7 & N  & 2014-10-22/05:17UT & S14E19 & 1.79$\pm$0.06 \\
18 & 12192 & X1.6 & N  & 2014-10-22/14:28UT & S14E13 & 2.13$\pm$0.05 \\
19 & 12192 & M1.1 & N  & 2014-10-23/09:50UT & S16E03 & 1.83$\pm$0.02 \\
20 & 12192 & M4.0 & Y  & 2014-10-24/07:48UT & S19W06 & 0.79$\pm$0.04 \\
21 & 12192 & X3.1 & N  & 2014-10-24/21:40UT & S16W21 & 1.15$\pm$0.08 \\
22 & 12673 & X2.2 & N  & 2017-09-06/09:10UT & S08W32 & 4.00$\pm$0.50 \\
\hline
\end{tabular}
\label{tab2}
\end{table*}

\subsection{Regions of interest}
\label{sect:pil}

In our previous work \citep{moraitis21} we considered the locations of the flare ribbons as the region of interest (ROI), as these correspond to the footprints of the structures that separate the erupting flux rope from the surrounding, confining field \citep{janvier13}. Their identification was done manually from inspection of the 1600~\r{A} channel of SDO's Atmospheric Imaging Assembly (AIA) instrument \citep{aiapaper}. This cannot be done in this work since the number of flares we are interested in is much larger, and also, a more automated process is desirable. We thus consider the magnetic PILs (MPILs) as the ROIs for this work. These can be identified in an automated fashion, following the steps described in \citet{schrijver07}. We use the same parameters as in that work and choose the strong-field threshold at 150~G and a dilation window of 3x3 pixels. Based on the original MPIL, we also calculate the region that results from the convolution of the MPIL with an area-normalized Gaussian with a full width at half maximum (FWHM) of 9 pixels. We denote the Gaussian MPIL as $W_\mathrm{MPIL}$ and use this in the following. For ARs 11158 and 12192 where the pixel size is twice as in the others ARs, the FWHM of the Gaussian is taken equal to 4.5 pixels and the dilation window to 2x2 pixels so that they correspond roughly to the same physical size as in the other ARs.

We note here that the choice of the MPILs as ROIs is very common when examining possible eruptivity indicators \citep[e.g., Sect.~3.2.5 of][and references therein]{guennou17}. Moreover, the flare ribbons that were examined in \citet{moraitis21} are usually located around the MPILs, and so the current choice is not that different. Furthermore, the 9-pixels FWHM of the Gaussian can be justified observationally. Taking the average ribbon area, $\sim 2\,10^{18}\,\mathrm{cm}^2$, from the catalogue of \citet{kazachenko17}, which comprises over 3000 above-C1.0-class solar flare ribbons observed between 2010-2016, we find that the average ribbon occupies $\sim 400$ pixels in a magnetogram of $1"$ pixel size. Further assuming that the ribbon shape is roughly a rectangle with ratio of sides 1:5, we find a length for the smaller side equal to $\sim 9$ pixels, consistent with our choice for the Gaussian.

\begin{figure*}[h]
\centering
\includegraphics[width=0.32\textwidth]{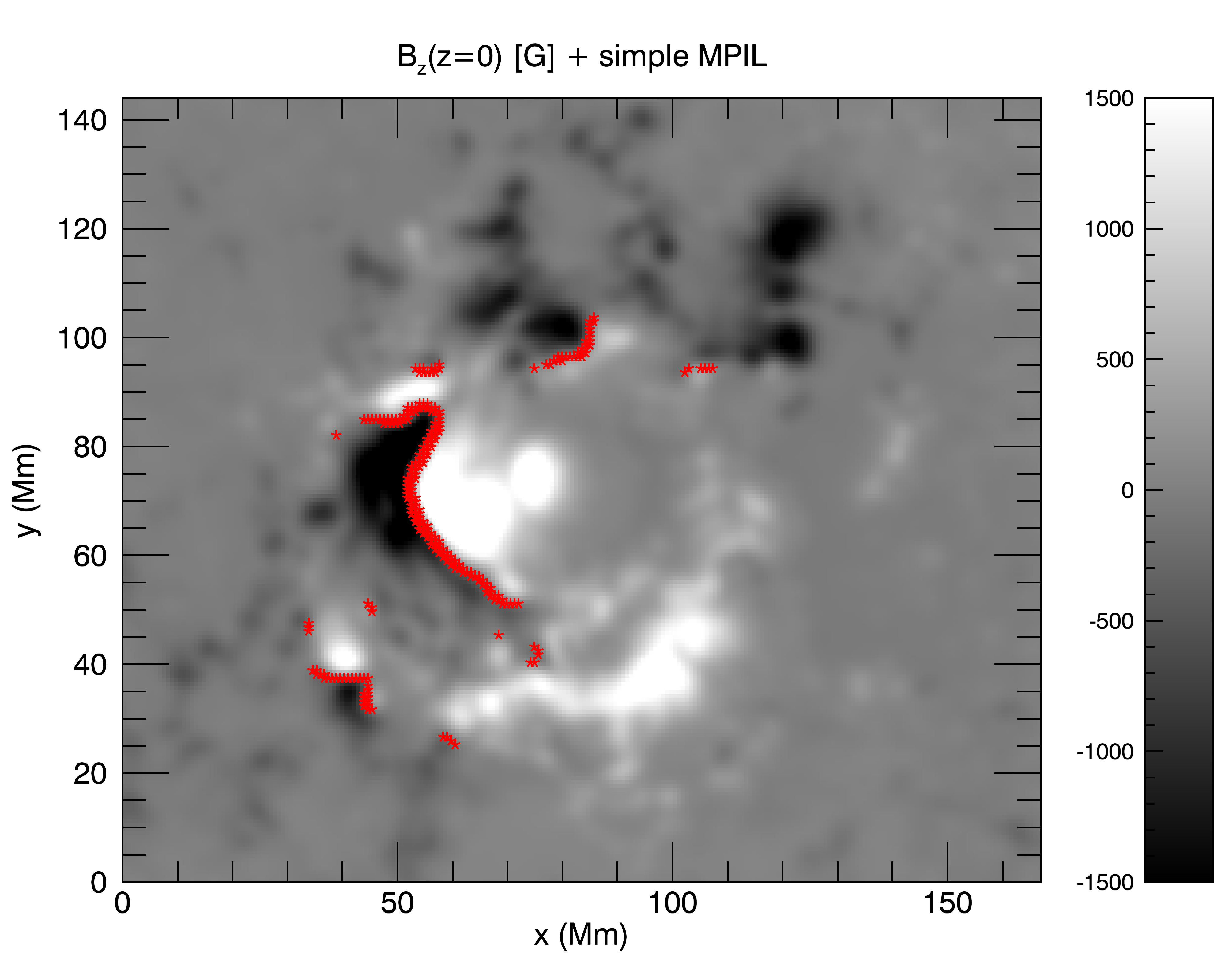}%
\includegraphics[width=0.32\textwidth]{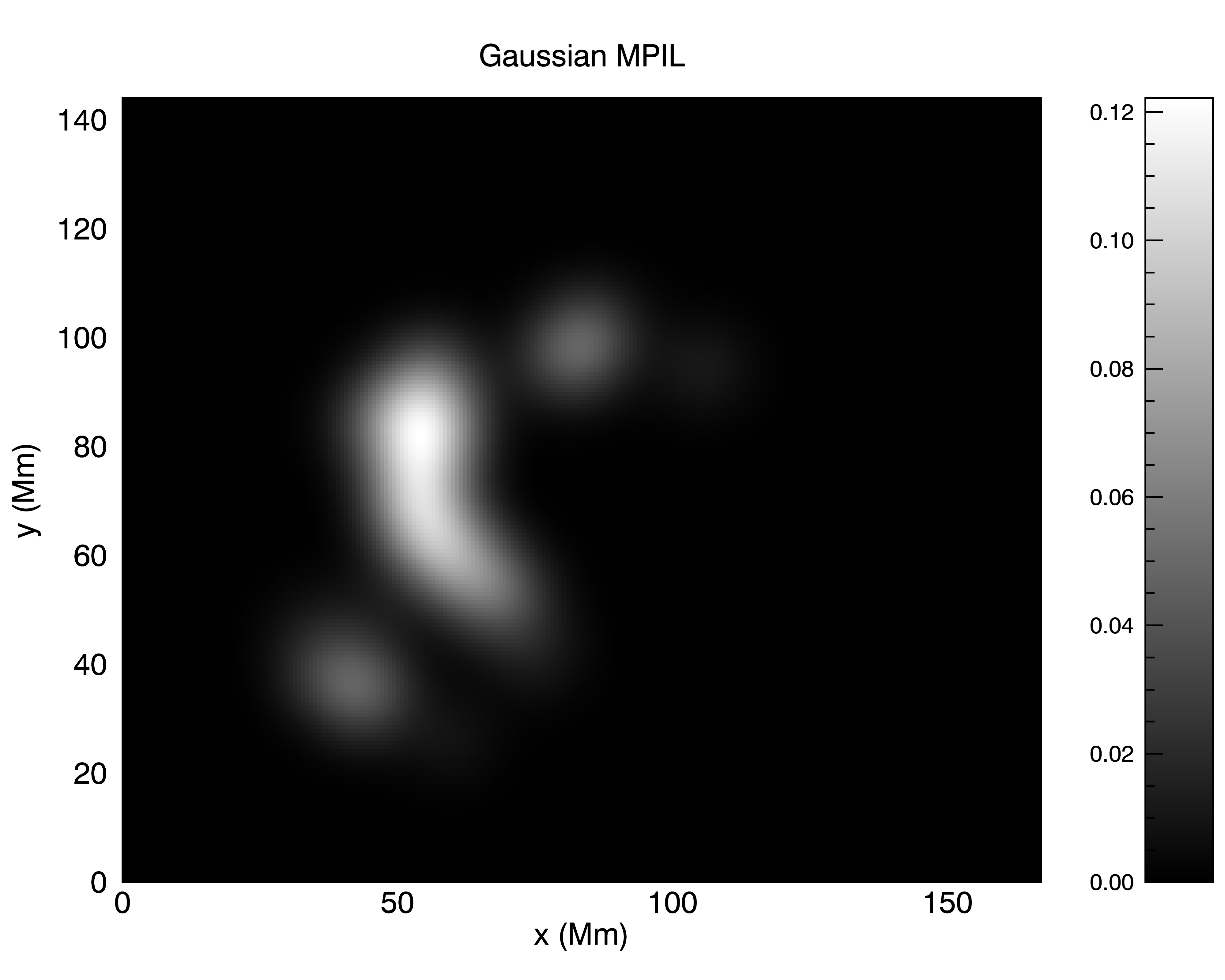}%
\includegraphics[width=0.32\textwidth]{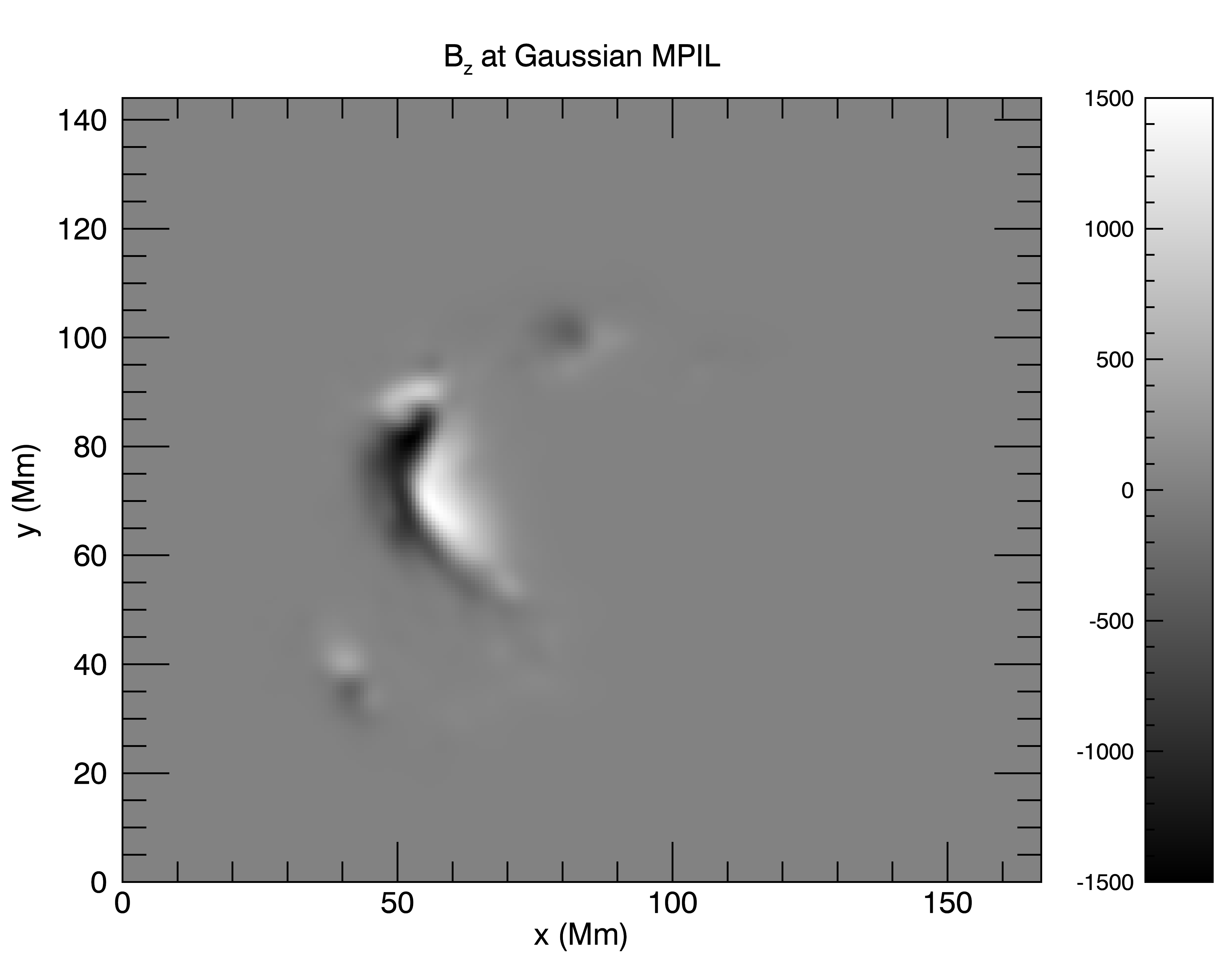}\\
\includegraphics[width=0.32\textwidth]{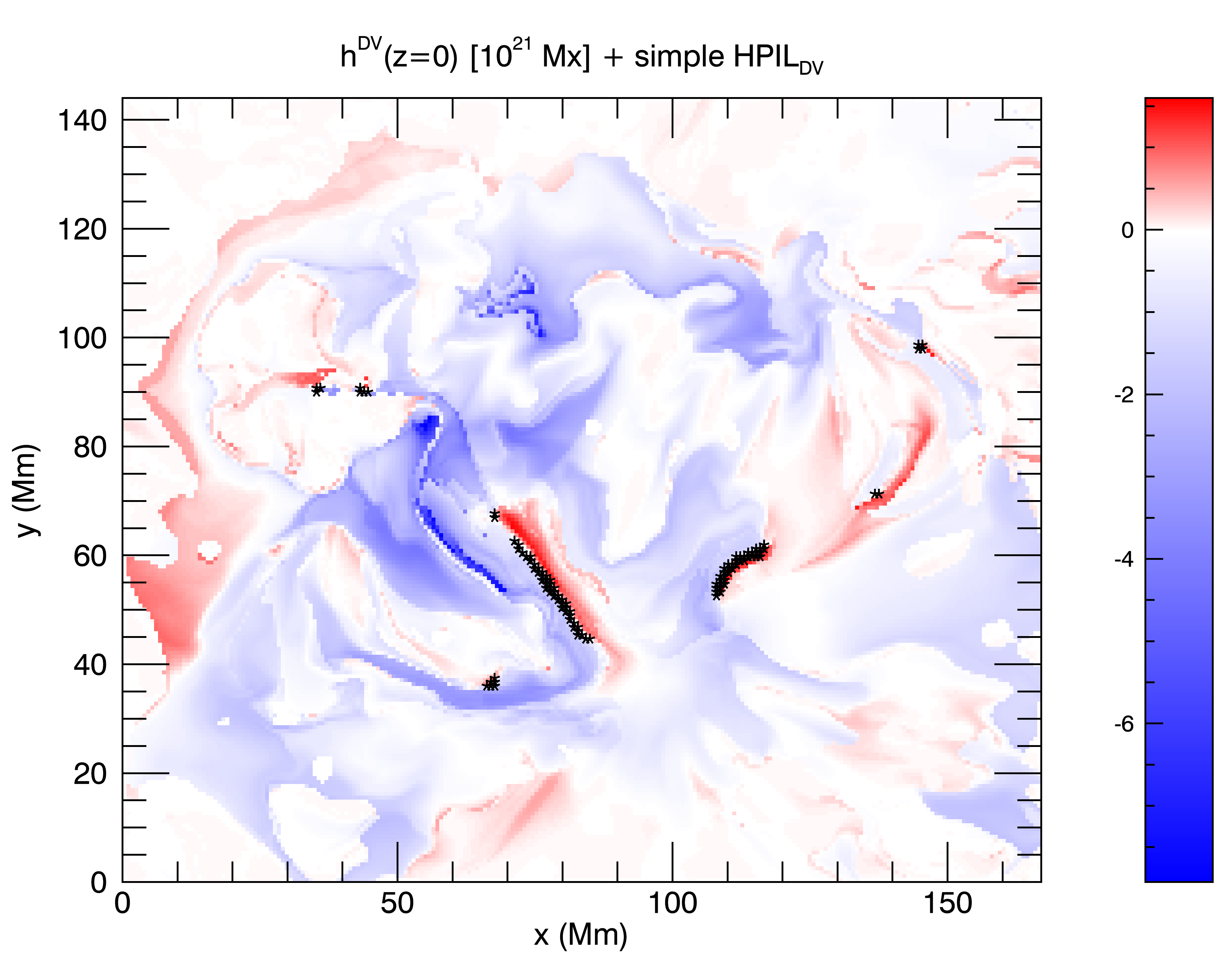}%
\includegraphics[width=0.32\textwidth]{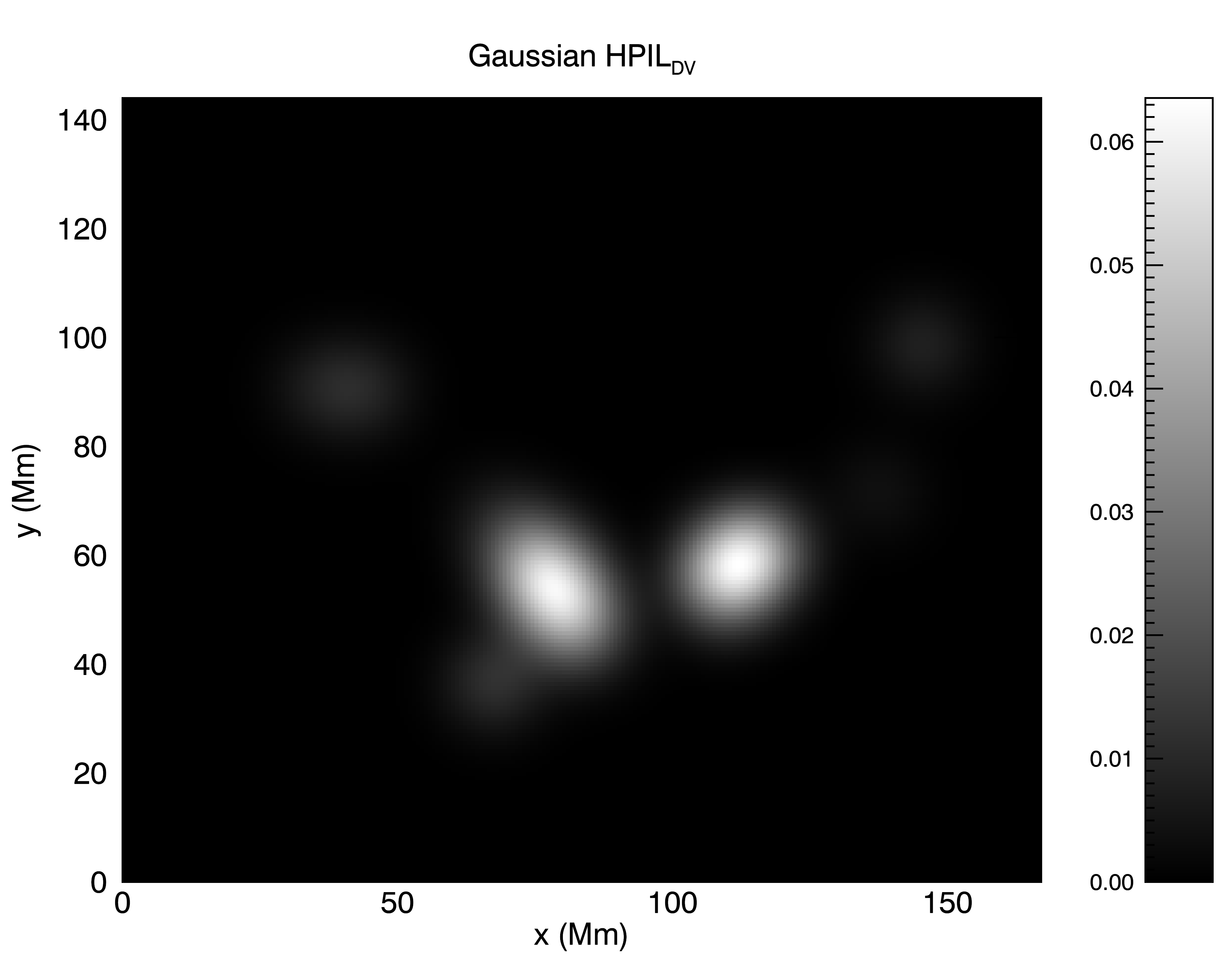}%
\includegraphics[width=0.32\textwidth]{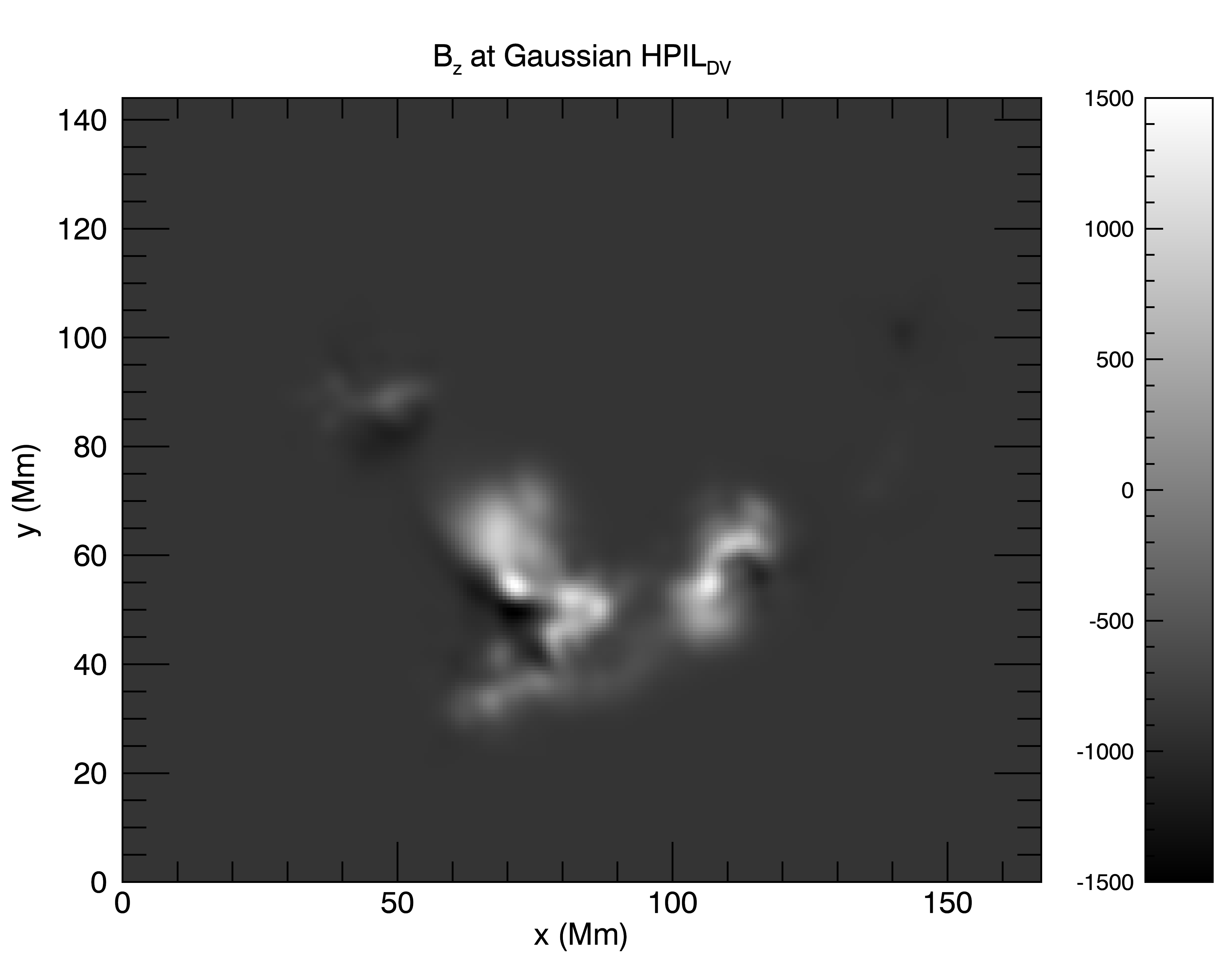}
\caption{Simple MPIL (red symbols) overlaid on the photospheric distribution of $B_z$ (left), respective Gaussian MPIL mask (middle), and the magnetogram after applying the Gaussian mask on it (right), for AR 12673 on 6 September 2017, 10:24 UT (top row). Similarly, in the bottom row, simple HPIL$_\mathrm{DV}$ (black symbols) on top of the photospheric distribution of RFLH in the DV gauge (left), respective Gaussian HPIL$_\mathrm{DV}$ mask (middle), and magnetogram after applying the Gaussian HPIL$_\mathrm{DV}$ mask on it (right).}
\label{pilfig}
\end{figure*}

An example of the morphology of the original and Gaussian MPILs for a snasphot of AR 12673 is shown at the top row of Fig.~\ref{pilfig}, along with the original, and the resulting magnetogram after applying the Gaussian mask on it. The MPIL in this case is a clear line, but in other ARs and/or snapshots the identified MPILs consist of very few points and cannot be considered as lines. We nevertheless choose to keep whatever is produced by the automated ROIs identification algorithm.

Apart from the MPILs we also determine the respective RFLH-based PILs, that is, the locations on the photosphere where RFLH changes sign. These helicity PILs (HPILs) are the locations where the handedness of the field changes. We consider separately the HPIL of RFLH in the DV gauge, which we denote as HPIL$_\mathrm{DV}$, and of RFLH in the BF gauge, HPIL$_\mathrm{BF}$. The HPILs are produced similarly to the MPILs, by taking the threshold of strong RFLH (in either gauge) at 10$\%$ of its absolute maximum value, and the same dilation window of 3x3 pixels, or 2x2 for ARs 11158 and 12192. This choice for the threshold corresponds roughly to the one made for the MPILs, as 150~G is on the order of 10$\%$ of the maximum value of $|B_z|$. Similarly with the MPIL, we define the Gaussian HPIL masks $W_\mathrm{HPIL_\mathrm{DV}}$ and $W_\mathrm{HPIL_\mathrm{BF}}$, and use them in the following. As done in the case of the MPILs, we perform no check on the resulting HPILs and accept whatever the algorithm produces. An example of the HPIL$_\mathrm{DV}$ for the case of AR 12673 is shown in the bottom row of Fig.~\ref{pilfig}. Regarding HPIL$_\mathrm{BF}$, this is similar to HPIL$_\mathrm{DV}$ in all cases, with only minor differences appearing between the two, and so is not depicted in Fig.~\ref{pilfig}.

In the specific AR and snapshot of Fig.~\ref{pilfig}, the two PILs, magnetic and helicity-based, are very different and have little to none common points. This is not always the case, as the morphology of the two PILs varies a lot among ARs and/or snapshots of the same AR, and there are cases when the magnetic and helicity PILs overlap to a higher degree. This is reasonable of course, as the two PILs, magnetic and helicity ones, have totally different physical meanings.

\subsection{Flare-related changes of relative helicities}
\label{sect:res2}

We now focus on intervals $\pm$1 hour around the peaks of the flares. More specifically, we identify the last point before the peak of each flare in the soft X-rays, as measured by the GOES satellites, and then consider five points (of 12-min cadence) before that point and five after it, so that the time profiles consist of 11 points, or $\sim 2$ hours, in total. We examine the evolution of various relative helicities around the peaks of the 22 flares:

\begin{itemize}
\item The relative helicity as computed from the volume method of Eq.~(\ref{helr}), $H_\mathrm{r}$.
\item The relative helicity as computed from the RFLH method in the DV gauge, $H_\mathrm{r}^\mathrm{DV}$, using Eqs.~(\ref{flhhel})-(\ref{flhdef2}).
\item The relative helicity as computed from the RFLH method in the BF gauge, $H_\mathrm{r}^\mathrm{BF}$, using Eqs.~(\ref{flhyp1})-(\ref{flhyp2}).
\end{itemize}
Apart from these, we also define the helicities contained in the (Gaussian) MPIL as computed from RFLH in the two gauges, $H_\mathrm{MPIL}^\mathrm{DV}$ and $H_\mathrm{MPIL}^\mathrm{BF}$. Their respective expressions are given by the restriction of Eq.~(\ref{flhhel}) on the location of the MPIL, which is achieved through the Gaussian weighting mask $W_\mathrm{MPIL}$, namely
\begin{equation}
H_\mathrm{MPIL}^\mathrm{DV}=\int h_\mathrm{r}^\mathrm{DV}\,W_\mathrm{MPIL}\,{\rm d}\Phi,
\label{hmpil}
\end{equation}
and
\begin{equation}
H_\mathrm{MPIL}^\mathrm{BF}=\int h_\mathrm{r}^\mathrm{BF}\,W_\mathrm{MPIL}\,{\rm d}\Phi.
\label{hmpil2}
\end{equation}
Similarly, we define the helicity that is contained in the (Gaussian) HPIL$_\mathrm{DV}$ as computed by RFLH in the DV gauge as
\begin{equation}
H_\mathrm{HPIL}^\mathrm{DV}=\int h_\mathrm{r}^\mathrm{DV}\,W_\mathrm{HPIL_\mathrm{DV}}\,{\rm d}\Phi,
\label{hhpila}
\end{equation}
and in the HPIL$_\mathrm{BF}$ as computed by RFLH in the BF gauge as
\begin{equation}
H_\mathrm{HPIL}^\mathrm{BF}=\int h_\mathrm{r}^\mathrm{BF}\,W_\mathrm{HPIL_\mathrm{DV}}\,{\rm d}\Phi.
\label{hhpilb}
\end{equation}
We should note here that the helicities of Eqs.~(\ref{hmpil})-(\ref{hhpilb}) are not strictly relative helicities, as they do not involve the potential field of the respective ROI but the one that corresponds to the whole volume. They can be considered as the part of the relative helicity that is contained in the respective PIL. In the following we thus use both the terms PIL helicity and PIL relative helicity for them. We also remind that apart from $H_\mathrm{r}$ all other helicities are gauge-dependent and therefore come in DV-BF pairs.

\begin{figure*}[h]
\centering
\includegraphics[width=0.32\textwidth]{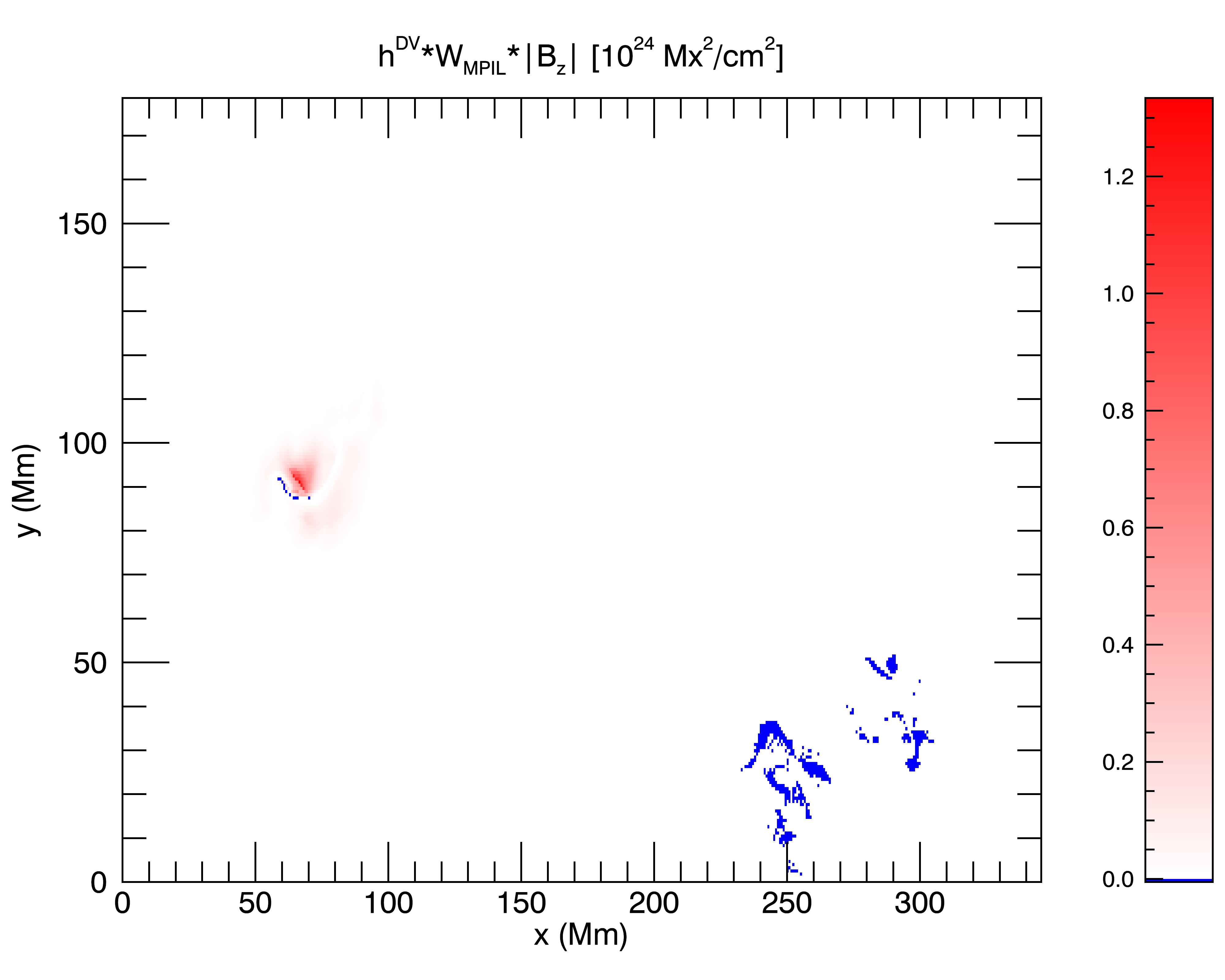}%
\includegraphics[width=0.32\textwidth]{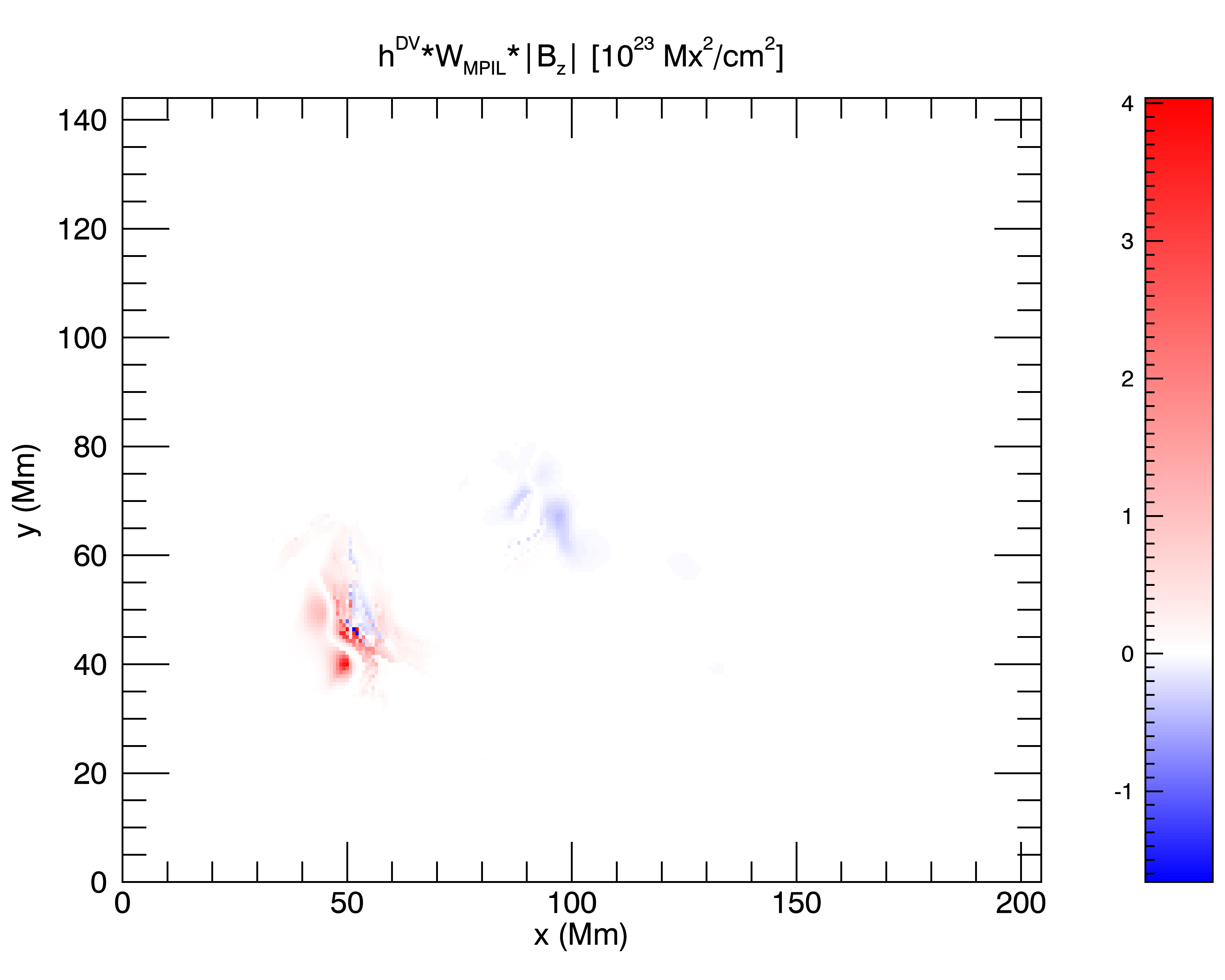}%
\includegraphics[width=0.32\textwidth]{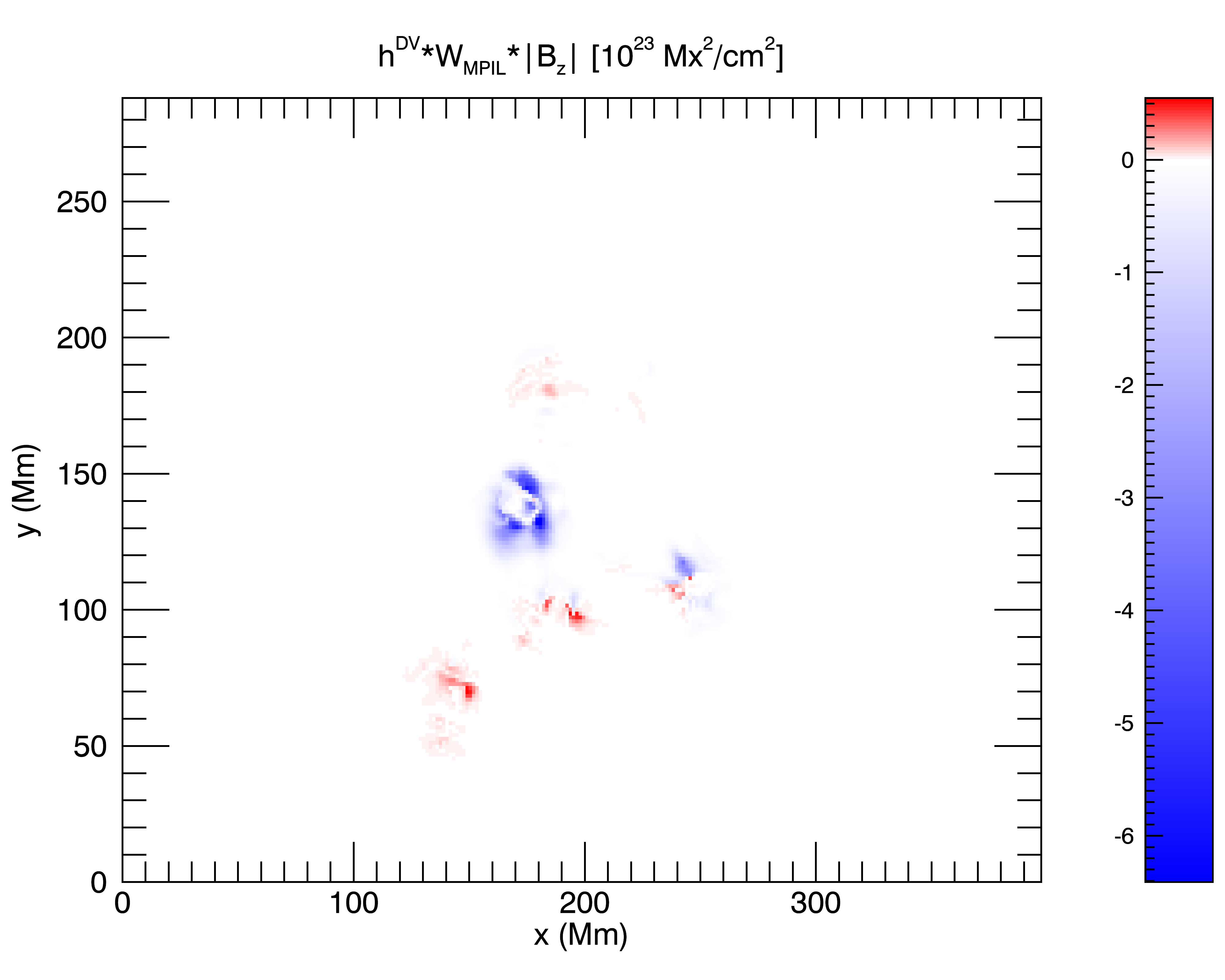}\\
\includegraphics[width=0.32\textwidth]{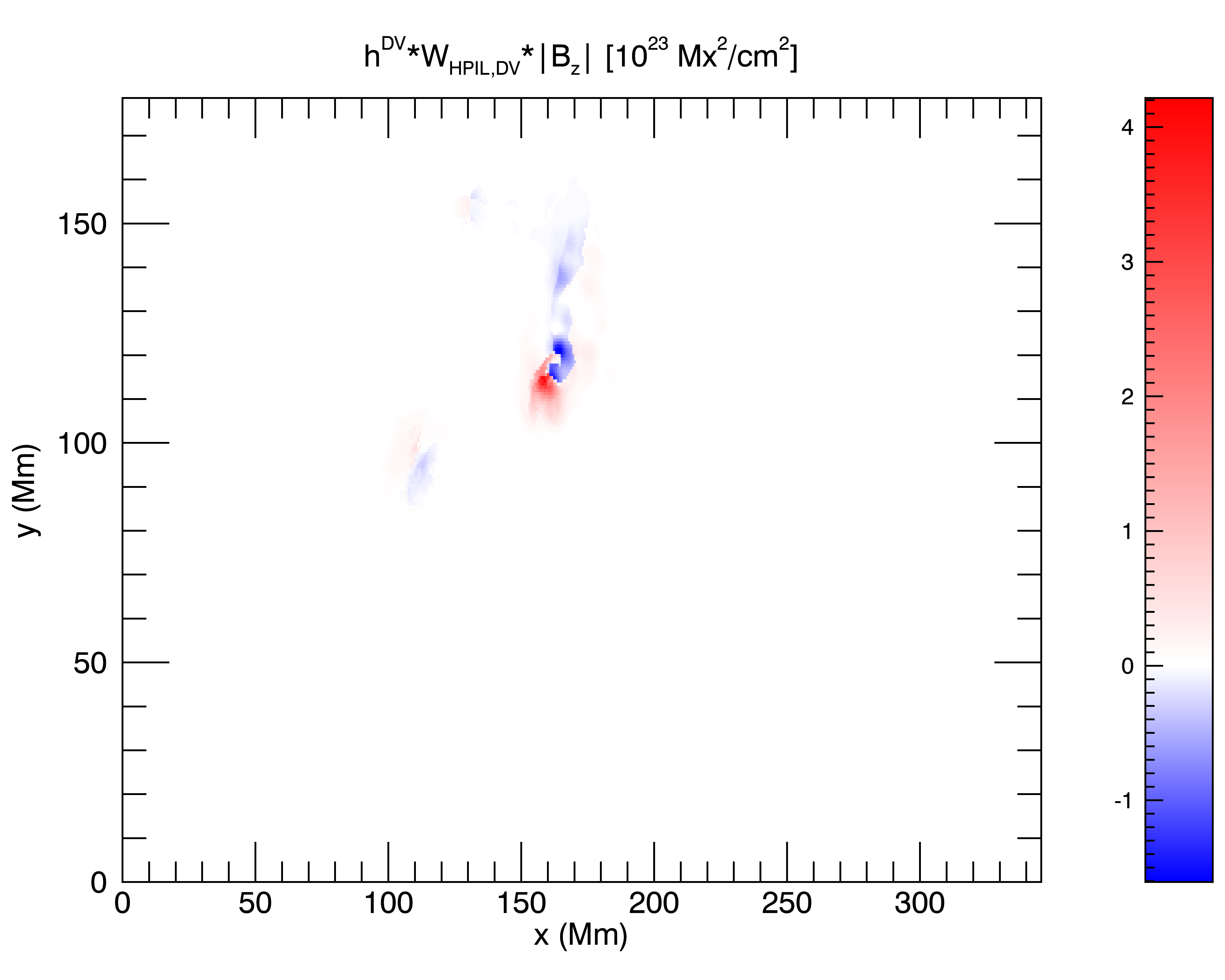}%
\includegraphics[width=0.32\textwidth]{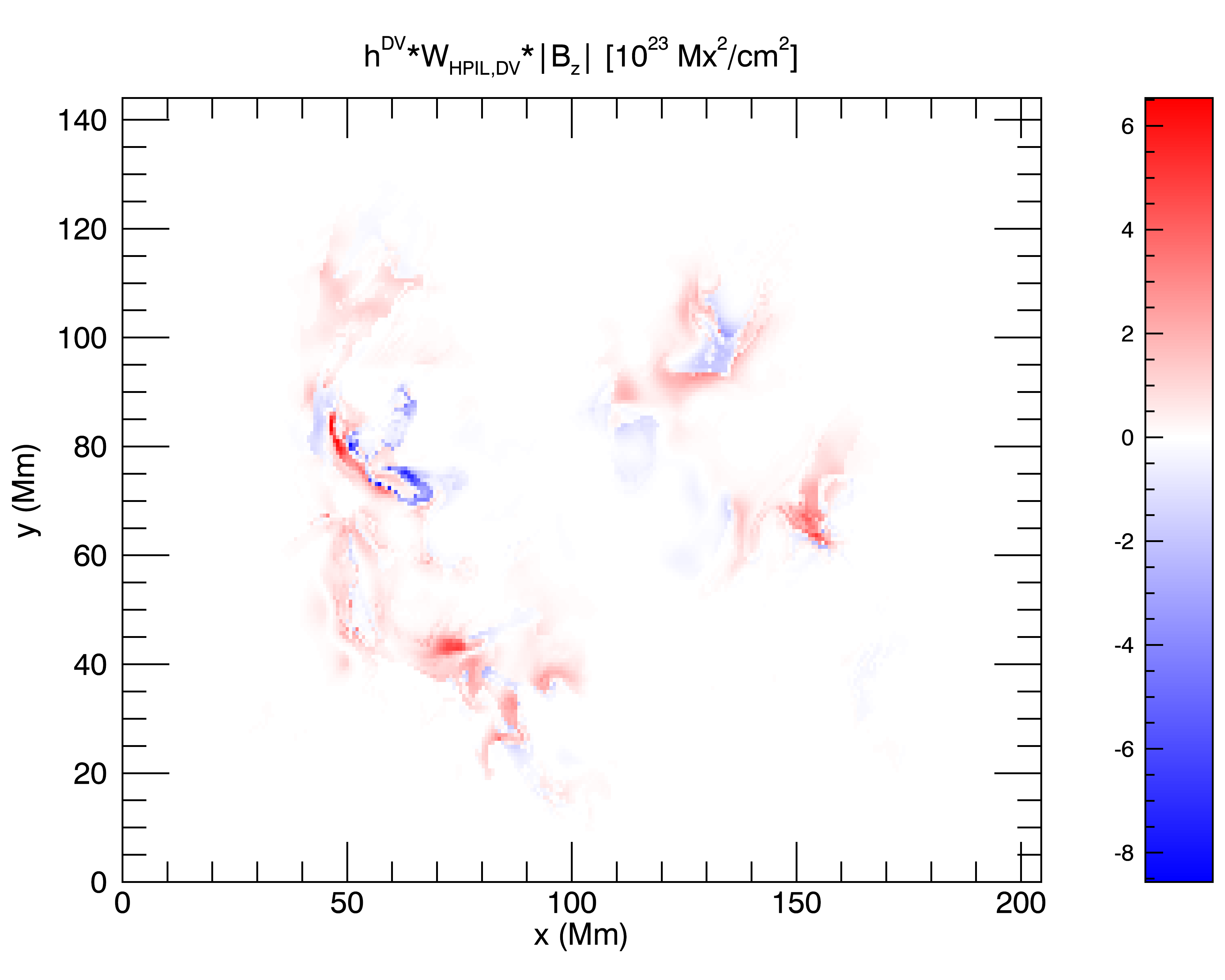}%
\includegraphics[width=0.32\textwidth]{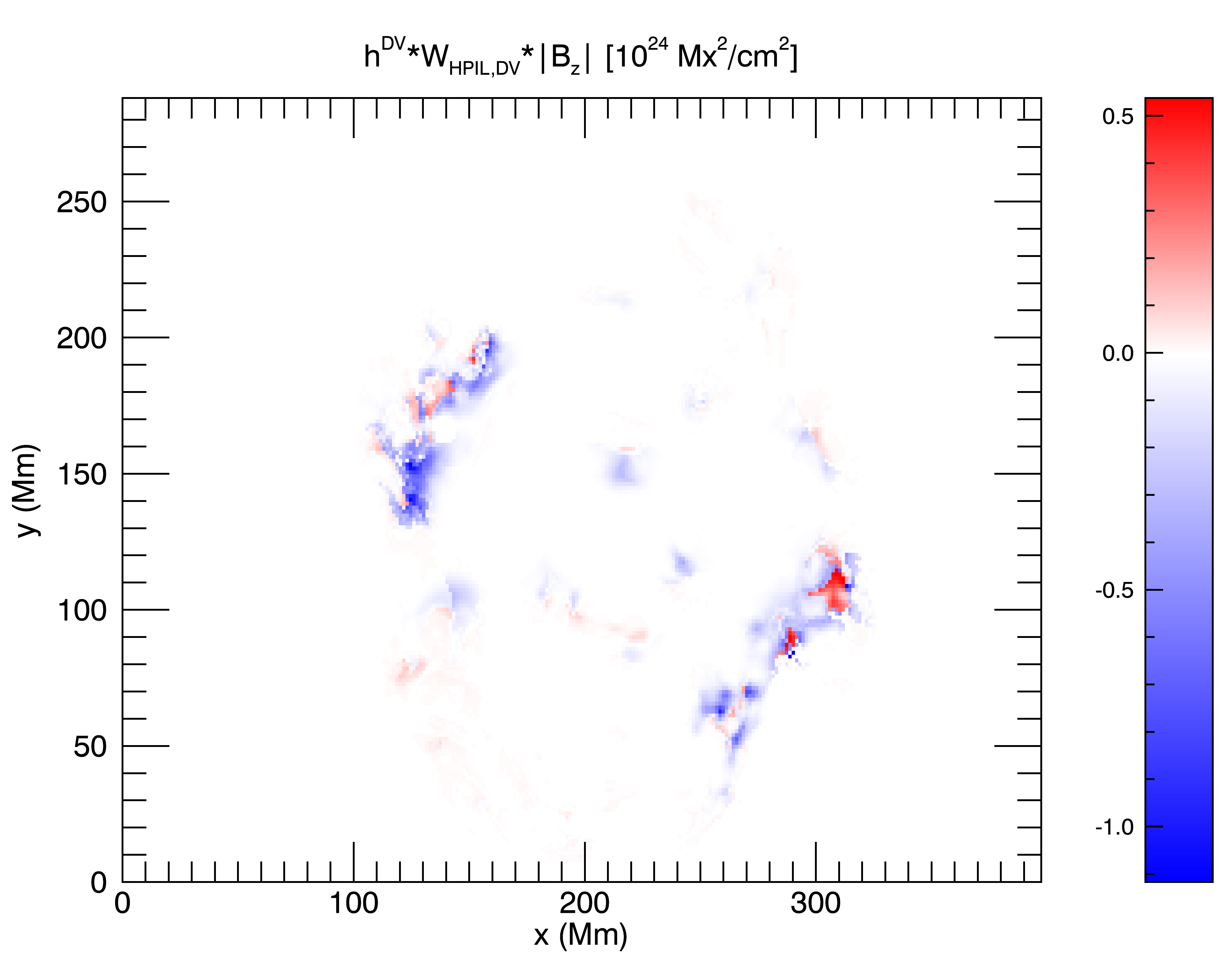}
\caption{Relative helicity distribution around the MPIL (top row) and the HPIL$_\mathrm{DV}$ (bottom row) for AR 11520 at 12 July 2012, 16:48 UT (left column), AR 11890 at 7 November 2013, 03:36 UT (middle column), and AR 12192 at 22 October 2014, 14:23 UT (right column).}
\label{pilfig2}
\end{figure*}

Before looking at the various helicities we further examine the distribution of RFLH around the two PILs, using the DV gauge as an example. We show in Fig.~\ref{pilfig2} the distribution of the integrands of Eqs.~(\ref{hmpil}) and (\ref{hhpila}) multiplied by $|B_z|$, so that summing the pixels in the shown images to directly result in the respective PIL helicity. Three ARs are depicted in Fig.~\ref{pilfig2} where the snapshot for each AR corresponds to the time right before the peak of individual flares. We notice that the distribution of helicity around the MPIL mostly follows the dominant helicity sign of the specific AR. This is also supported by Fig.~\ref{flhmorph}, while evidence for it has been seen previously \citep{moraitis19,moraitis21}. A physical explanation for this association comes from the standard 3D flare model \citep{janvier14}, according to which, the highest twist during a flare is located at the footpoints of a flux rope, which lie on either side of the MPIL. We therefore expect that the values of RFLH are the highest at the MPIL, and that their sign is consistent with that of $H_\mathrm{r}$.

For the distribution of helicity around the HPIL$_\mathrm{DV}$, which exhibits both signs of helicity on each side of the HPIL$_\mathrm{DV}$ by definition, we note on the other hand that it is more or less balanced. One could therefore expect that helicity cancels out and is very weak there, but Fig.~\ref{pilfig2} shows that there are cases when the values of RFLH around the HPIL are more important than those found around the MPIL. This could be explained by the higher values of RFLH around the HPIL and/or the larger number of pixels this occupies. Nevertheless, since both signs of RFLH are found around the HPIL, the respective helicity is expected to show more complicated behaviour than the MPIL one.

\begin{figure*}[h]
\centering
\includegraphics[width=0.48\textwidth]{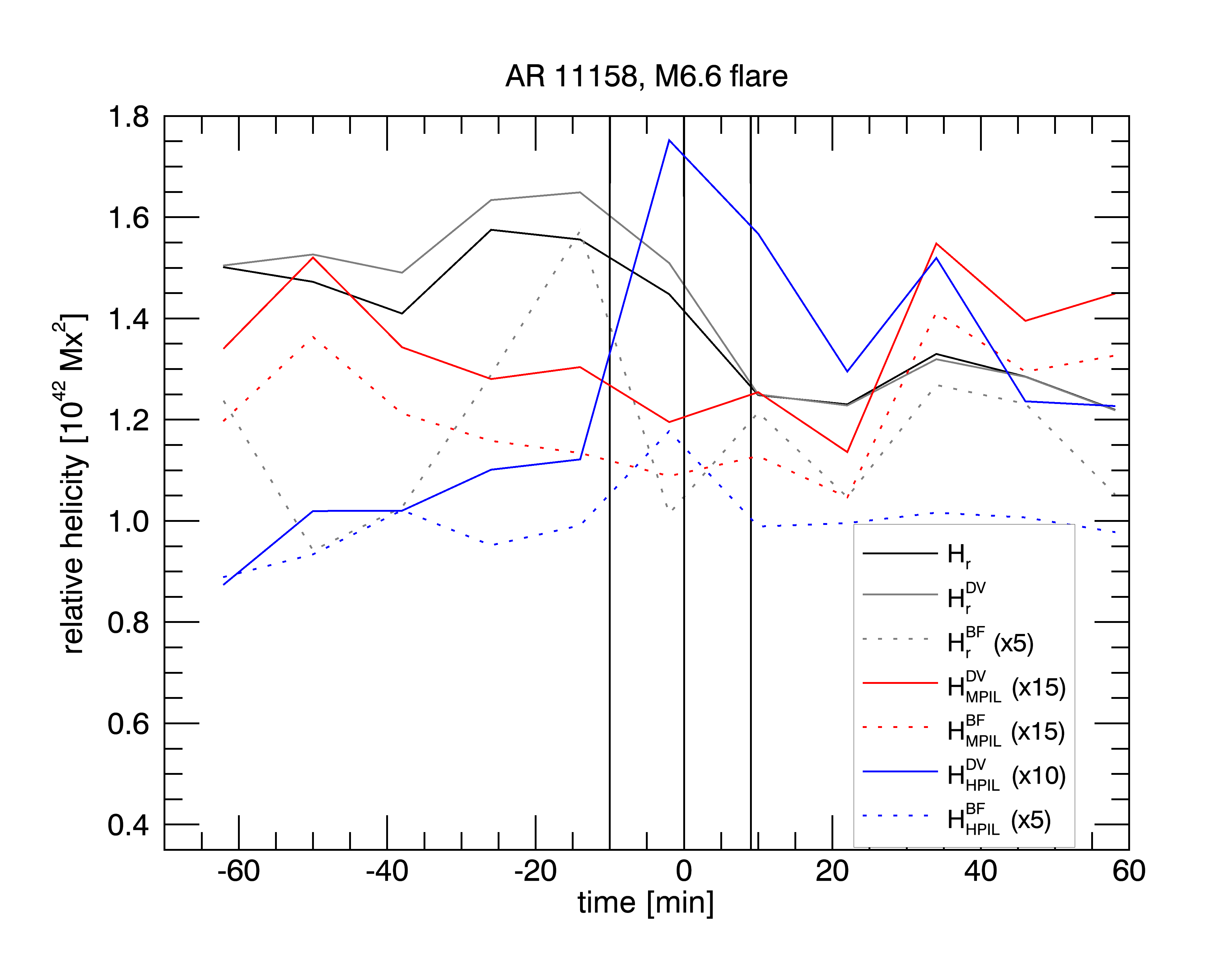}%
\includegraphics[width=0.48\textwidth]{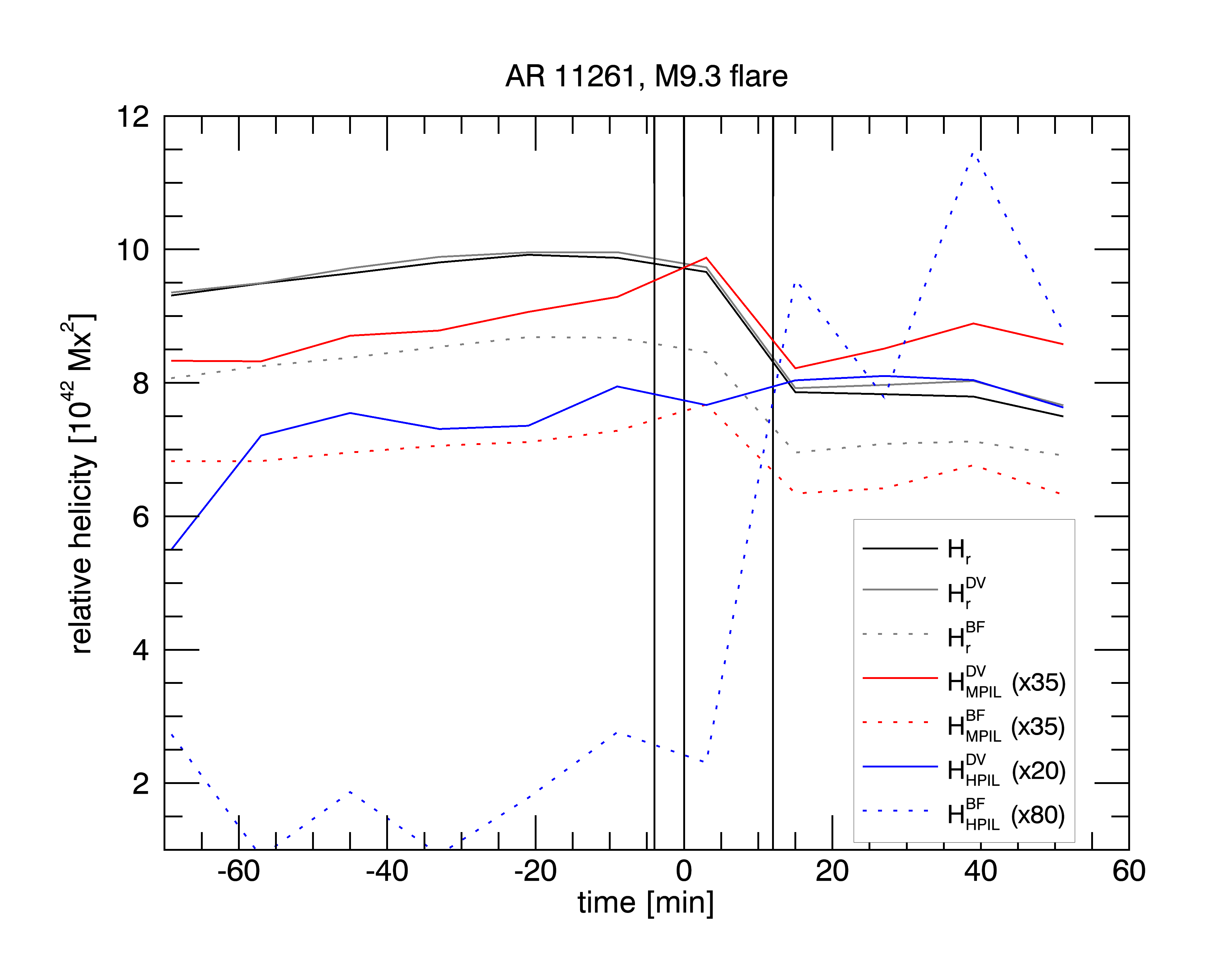}\\
\includegraphics[width=0.48\textwidth]{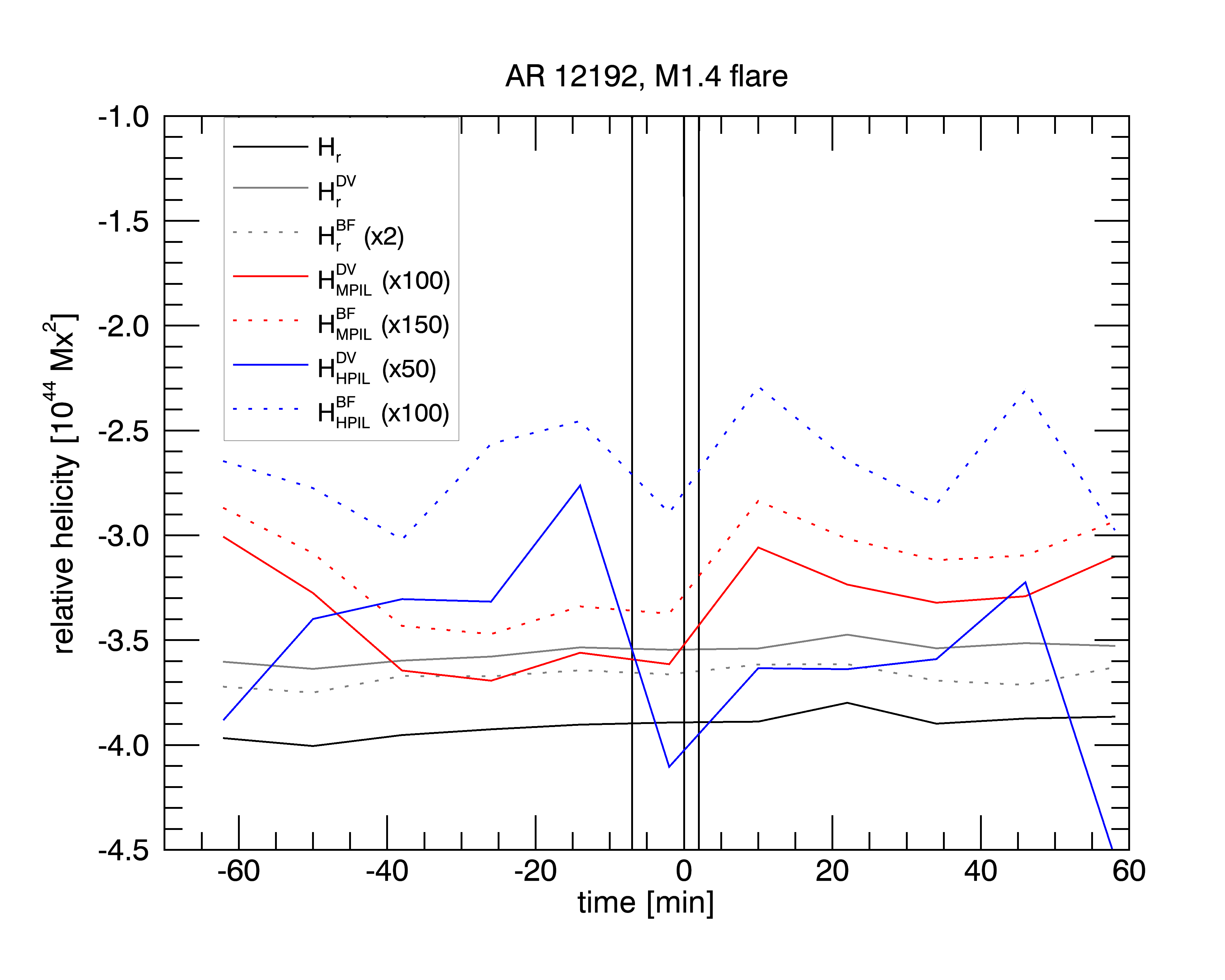}%
\includegraphics[width=0.48\textwidth]{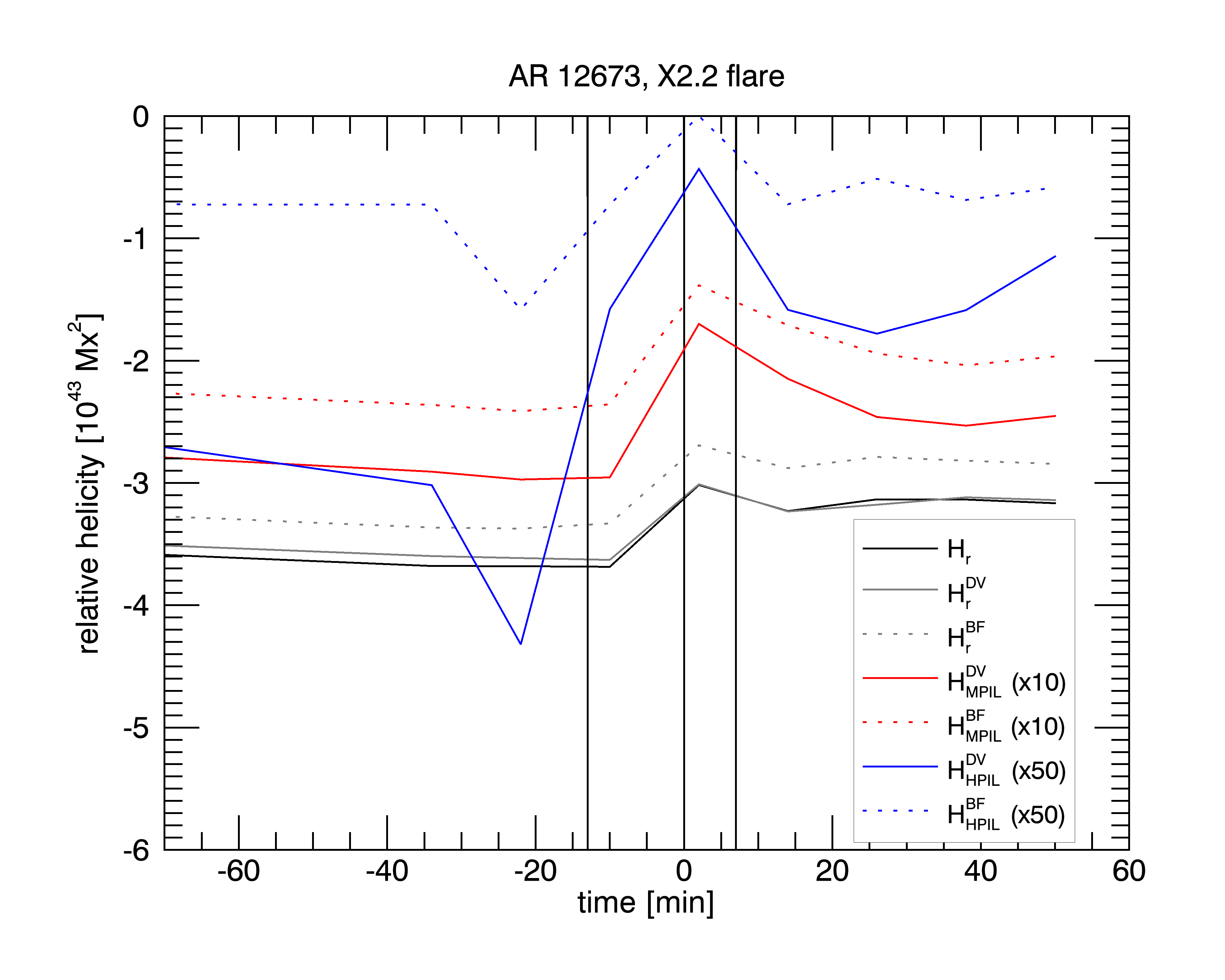}
\caption{Evolution of various relative helicities around four of the examined flares. Two of them are eruptive, in ARs 11158 and 11261 (top row), and two confined, in ARs 12192 and 12673 (bottom row). Shown are the relative helicities as computed from the volume method (black solid line), the RFLH in the DV gauge (grey solid line), and the RFLH in the BF gauge (grey dotted line). Also shown are the relative helicities at the MPIL (red lines; solid: DV, dotted: BF), and at the two HPILs (blue lines; solid: DV, dotted: BF). The vertical black lines denote the start, peak, and end times of the flares (from left to right), after the curves have been shifted so that the peak time corresponds to the zero time. The multiplication factors shown in the legends are used for better visualization of the respective curves.}
\label{dataplot2}
\end{figure*}

In Fig.~\ref{dataplot2} we show the evolution of all the relative helicities for four of the flares from four different ARs. In all the plots of Fig.~\ref{dataplot2} the volume curves serve as the ground truth which provide the most accurate values for relative helicity, as is deduced by the consistency between the various finite-volume computational methods \citep{valori16,thalmann21}. The grey curves are used as indicators of how well the respective methods reproduce the volume helicity. We note that the DV gauge is usually closer to the volume method than the BF gauge, DV is on average (for all flares) 15\% away from the volume curves while BF 47\%, although the differences vary from case to case and range between $(0\%,95\%)$ and $(6\%,96\%)$, in the DV and BF gauges respectively. This is something already noticed in \citet{moraitis21}. Although a possible reason for the disagreement between the helicities computed by a volume and an RFLH method is the large number of field lines that close within the examined volume \citep{moraitis19}, this is probably not the case here, and the BF gauge simply produces systematically smaller values. A more plausible explanation could be that the restriction of the RFLH computations on the photospheric boundary has different effect on the two gauges, leading to smaller BF values.

Regarding the MPIL relative helicities (red curves), we note that their time profiles show a more or less qualitative agreement with those of the volume curves. In all cases of Fig.~\ref{dataplot2}, the MPIL helicities exhibit a decrease during the flares, as the respective volume curves. Quantitavely, the MPIL helicities differ from the volume ones by a large factor of course, ranging in the interval $(1,400)$ for all the flares in the DV gauge (with a mean value of 90), and in $(2,600)$ for the BF one (with a mean value of 140), as the MPILs usually occupy very small areas. The agreement between the two gauges of the MPIL helicities in Fig.~\ref{dataplot2} (red solid and dotted lines) is generally good, although, it too, varies among cases.

The general agreement in the temporal evolution of the MPIL helicities with the ground-truth volume helicity is consistent with the standard 3D flare model \citep{janvier14}. A decrease of $|H_\mathrm{MPIL}|$ is related with a similar decrease of the dominant sign of helicity and the eventual small increase, or the appearance, of parasitic opposite helicity, something also evident in \citet[][Fig.~7]{moraitis21}.

The HPIL relative helicities' time profiles (blue curves) show mixed behaviour regarding their qualitative agreement with the respective volume curves. There are cases, such as for AR 12673 in Fig.~\ref{dataplot2}, where there is agreement, and others, such as for AR 11261, where the HPIL helicities show profiles of different trend compared to the volume method, increasing instead of decreasing. Overall, the HPIL helicity profiles agree with the volume ones much less than the MPIL helicities. Additionally, they exhibit larger disagreement between the two gauges (blue solid and dotted lines) compared to the MPIL helicities, as can be deduced from Fig.~\ref{dataplot2}.

The HPIL helicities are related with complex evolution on the sides of the AR which involves both signs of helicity, and thus may present different behaviours. As an example, a decrease of $H_\mathrm{HPIL}$ (in either gauge) may mean that there is a strong decrease of the dominant-sign helicity and a smaller decrease of the minority-sign helicity, or a decrease of the dominant-sign helicity and no change of the minority-sign helicity, or no change of dominant-sign and an increase of the minority-sign helicity, and so on. We also note that on average the HPIL helicities are larger in absolute value than the MPIL ones, although without universal behaviour; in some ARs the HPIL helicities are higher, while in others, the MPIL ones are. There is likely no universal trends of evolution of $H_\mathrm{HPIL}$ at, and around, solar flares.

\begin{figure}[h]
\centering
\includegraphics[width=0.48\textwidth]{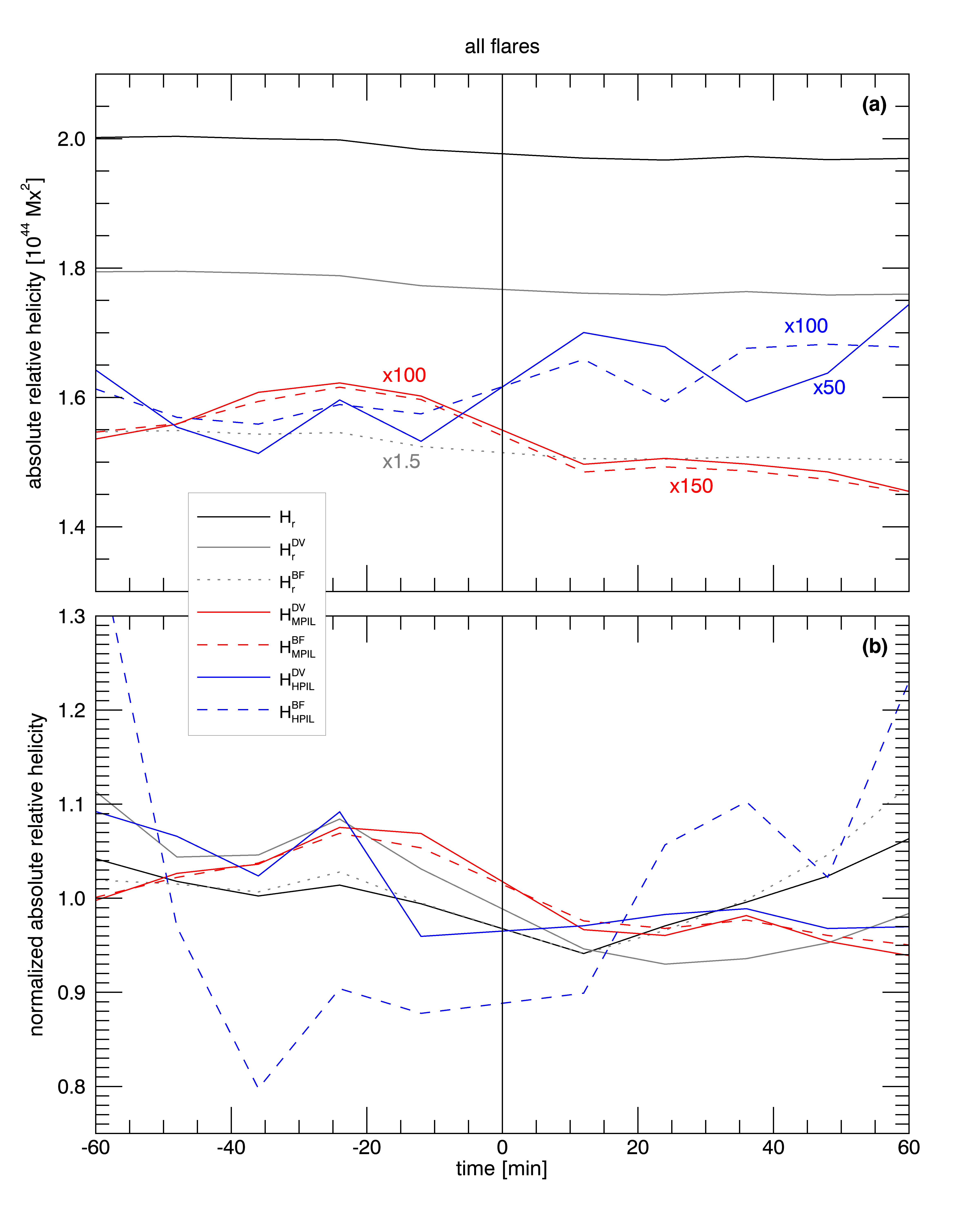}
\caption{Superposed epoch analysis of the absolute values of the relative helicities of Fig.~\ref{dataplot2} (top), and of their respective normalized-to-the-mean helicities (bottom), for all flares. The time of zero corresponds to the peak of the flares.}
\label{speplot}
\end{figure}

All these remarks are also depicted in the superposed epoch (SPE) analysis of the various relative helicity profiles of all flares, that is shown in Fig.~\ref{speplot}. By SPE we mean that each curve of Fig.~\ref{speplot} is the average of the respective individual profiles for all 22 flares, after the latter have been interpolated to the times $t_i= i\times(12\,\mathrm{min})$, to emulate the cadence of HMI observations. The non-zero integer $i\in [-5,5]$, and $t=0$ corresponds to the peak of the flares. The large helicity values for $H_\mathrm{r}$ in Fig.~\ref{speplot}a are mostly due to AR 12192, as they were more than a factor of 5 larger than in all other ARs. This large range of the individual helicities, in combination with the relatively small number of flares, is also the reason for the large errors (from averaging) of each curve, on the order of 20\%, which are thus not shown for the shake of clarity.

In order to quantify the helicity changes during the flares we use two quantities. Let $f$ stand for any of the helicity profiles and $f_i$ be its values at the ten interpolated times $t_i$. We then define the average relative change of $f$ between the pre-, and the post-flare values as
\begin{equation}
\Delta f=\frac{\sum_{i>0} f_i-\sum_{i<0} f_i}{\sum_{i<0} f_i},
\end{equation} 
and the instantaneous relative change as
\begin{equation}
\mathrm{D}f=\frac{f_1-f_{-1}}{f_{-1}}.
\end{equation}
The volume and RFLH helicities show marginal decreases during the flares, $\sim\,2\%$. The MPIL helicities show the more pronounced decreases during the flares compared to all other helicities; their relative average decrease is $\sim\,6.2\%$ for the DV gauge and $\sim\,6.6\%$ for the BF, with similar numbers for the instantaneous changes. Moreover, these decreases seem to start $\sim 20$ minutes before the peak of the flares. On the other hand, the HPIL helicities show a strong average increase during the flares, of $\sim\,6.6\%$ for the DV gauge and $\sim\,4.9\%$ for the BF, with even larger values for the instantaneous changes. All these results are summarized in Table~\ref{tab3}.

We repeat the above SPE analysis after having normalized each of the seven helicity profiles of Fig.~\ref{speplot}a by their mean values during each flare separately, and show the results in Fig.~\ref{speplot}b. This normalization scales the individual profiles around the value of 1 and reduces considerably the range between them, and as a result, the SPE averaging errors are decreased to values $\sim 5\%$. The errors are suppressed in Fig.~\ref{speplot}b as well, for reasons of clarity. The corresponding relative changes of helicity from this analysis are reported in Table~\ref{tab3}, next to the respective unnormalized ones. It is evident that all, but the two HPIL helicities, experience beyond-errors instantaneous decreases during the flares. The two MPIL helicities and $H_\mathrm{r}^\mathrm{DV}$ decrease beyond errors also on average. The unnormalized HPIL helicities profiles are incoherent, with both increasing and decreasing intervals, which is also depicted in the values of their relative changes during the flares, that are shown in Table~\ref{tab3}.

\begin{table}[ht]
\caption{Average ($\Delta f$), and instantaneous relative changes ($\mathrm{D}f$) during the flares of the SPE helicity profiles of Fig.~\ref{speplot}, original (left) and normalized versions (right).}
\centering
\resizebox{0.48\textwidth}{!}{
\begin{tabular}{ccc|ccc}
\hline
$f$ & $\Delta f (\%)$ & $\mathrm{D}f (\%)$ & $f$ & $\Delta f (\%)$ & $\mathrm{D}f (\%)$ \\
\hline
$H_\mathrm{r}$ & -1.4 & -0.7 & norm. $H_\mathrm{r}$ & -1.5 & -5.3 \\
$H_\mathrm{r}^\mathrm{DV}$ & -1.6 & -0.7 & norm. $H_\mathrm{r}^\mathrm{DV}$ & -10.7 & -8.2 \\
$H_\mathrm{r}^\mathrm{BF}$ & -2.4 & -1.2 & norm. $H_\mathrm{r}^\mathrm{BF}$ & 0.13 & -5.4 \\
$H_\mathrm{MPIL}^\mathrm{DV}$ & -6.2 & -6.6 & norm. $H_\mathrm{MPIL}^\mathrm{DV}$ & -7.7 & -9.6 \\
$H_\mathrm{MPIL}^\mathrm{BF}$ & -6.6 & -7.0 & norm. $H_\mathrm{MPIL}^\mathrm{BF}$ & -6.8 & -7.4 \\
$H_\mathrm{HPIL}^\mathrm{DV}$ & 6.6 & 11.0 & norm. $H_\mathrm{HPIL}^\mathrm{DV}$ & -6.7 & 1.2 \\
$H_\mathrm{HPIL}^\mathrm{BF}$ & 4.9 & 5.4 & norm. $H_\mathrm{HPIL}^\mathrm{BF}$ & 7.6 & 2.4 \\
\hline
\end{tabular}
}
\label{tab3}
\end{table}

\subsection{Relation with eruptivity}
\label{sect:res3}

We know from the previous Section, and especially from Figs.~\ref{dataplot2} and \ref{speplot}, that the relative helicity contained in the MPIL has a similar time profile around the studied flares as the ground-truth volume helicity, although with a much smaller magnitude. Here, we want to explore whether this PIL helicity has any relation with solar eruptivity.

A well-known, PIL-deduced parameter that is related with eruptivity is the $R$ parameter \citep{schrijver07}. This is simply the magnetic flux contained in the MPIL, $\Phi_\mathrm{MPIL}$, in units of magnetic field strength, that is, $R=\Phi_\mathrm{MPIL}/\lambda^2$ where $\lambda$ is the pixel size in physical units. ARs with large values of this parameter, $R\gtrsim10^5$~G, have high likelihoods for producing flares stronger than M1.0.

\begin{figure}[h]
\centering
\includegraphics[width=0.46\textwidth]{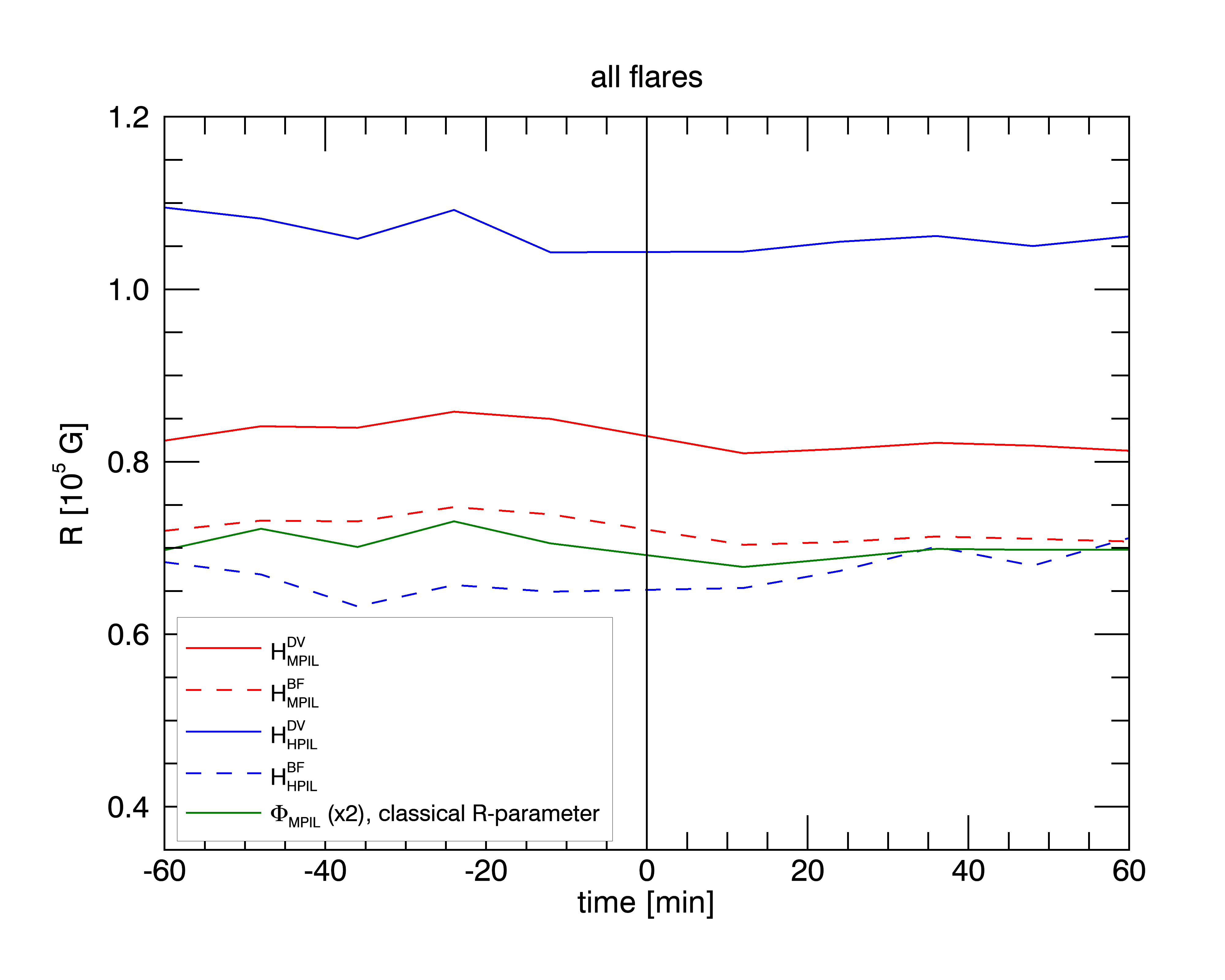}
\caption{Superposed epoch plot of the various $R$-parameters, for all the flares. Red and blue curves correspond to the $R_\mathrm{H}$ parameters thar are derived using $H_\mathrm{MPIL}^\mathrm{DV/BF}$ and $H_\mathrm{HPIL}^\mathrm{DV/BF}$, respectively (solid lines: DV gauge, dotted lines: BF gauge), while the green curve corresponds to the original $R$ parameter of \citet{schrijver07}. The zero of time corresponds to the peak of the flares.}
\label{speplot2}
\end{figure}

Using the PIL relative helicities of Sect.~\ref{sect:res2} we can define similar helicity-based $R$-parameters, through the relation
\begin{equation}
R_\mathrm{H}=\frac{H_\mathrm{PIL}^{1/2}}{\lambda^2},
\label{rh}
\end{equation}
where $H_\mathrm{PIL}$ can be any of the four helicities $H_\mathrm{MPIL}^\mathrm{DV}$, $H_\mathrm{MPIL}^\mathrm{BF}$, $H_\mathrm{HPIL}^\mathrm{DV}$, or $H_\mathrm{HPIL}^\mathrm{BF}$. The superposed epoch plot of the various $R_\mathrm{H}$ parameters along with the original parameter that is based on magnetic flux, is shown in Fig.~\ref{speplot2}. The MPIL $R_\mathrm{H}$'s (red curves) reaffirm the behaviour of the respective helicities in Fig.~\ref{speplot}, with a decrease of their average pre-flare values during the flares, of $\sim\,3.2\%$ for the DV gauge and $\sim\,3.5\%$ for the BF gauge. The respective $R$ profile (green curve) exhibits a milder decrease, $\sim\,2.7\%$, and is also smaller by a factor of two compared to the MPIL $R_\mathrm{H}$'s. The HPIL $R_\mathrm{H}$ parameters on the other hand (blue curves), exhibit mixed behaviour, with an average decrease during flares, of $\sim\,1.8\%$ for the DV gauge, and an average increase of $\sim\,3.9\%$ for the BF gauge.

We also note that by using the relative helicities computed by the volume and the RFLH methods, $H_\mathrm{r}$, $H_\mathrm{r}^\mathrm{DV/BF}$, in Eq.~(\ref{rh}) leads to much higher $R_\mathrm{H}$ values compared to those produced by the PIL ones, by almost an order of magnitude, as it is expected by the respective two-orders-of-magnitude-higher helicity budgets. The corresponding profiles however are much milder compared to those shown in Fig.~\ref{speplot2}, showing marginal decreases during flares of less than $1.5\%$, in agreement with the results of Table~\ref{tab3}.

\begin{figure}[h]
\centering
\includegraphics[width=0.48\textwidth]{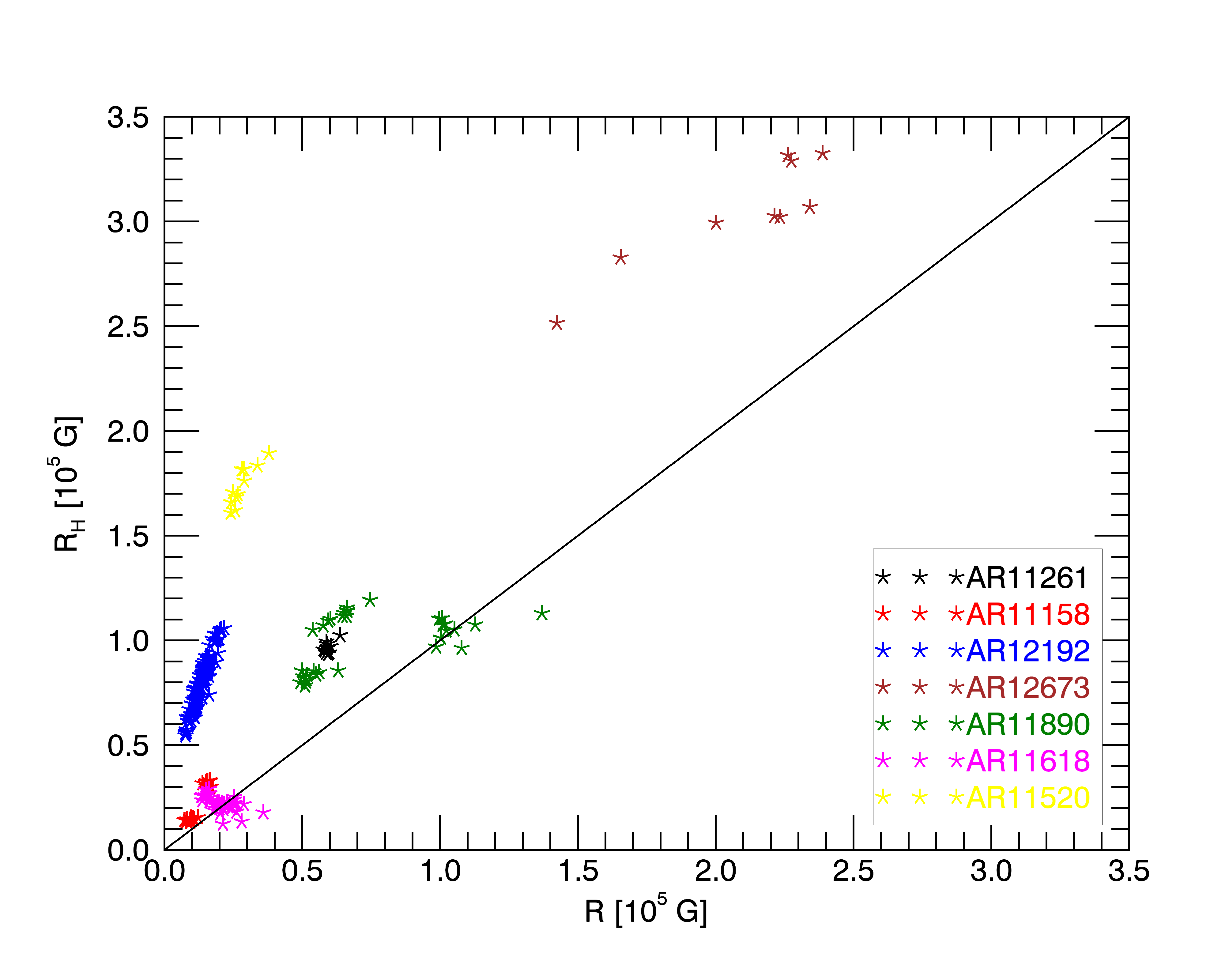}
\caption{Scatter plot of the values of $R_\mathrm{H}$ for the case of $H_\mathrm{MPIL}^\mathrm{DV}$ versus the respective $R$ values, for the 22 examined flares, coloured according to the specific AR. The two sets exhibit a linear correlation coefficient of 0.79. The solid line corresponds to the relation $R=R_\mathrm{H}$.}
\label{boxplot}
\end{figure}

It seems therefore that the $R_\mathrm{H}$ parameters which are based on the MPIL helicities have the highest potential to be used as eruptivity indicators. To further stress on that, we compare in Fig.~\ref{boxplot} the values of $R_\mathrm{H}$ for the case of $H_\mathrm{MPIL}^\mathrm{DV}$ at the times $\pm$~1h around the 22 studied flares with the corresponding values for $R$. We could equally use any of the two gauges for the MPIL helicity, but we choose to show the results for the DV gauge. We note that $R_\mathrm{H}$ exhibits similar values to $R$ in three ARs, and even larger values in the other four ARs. This means that $R_\mathrm{H}$ can be as good as, and in some cases, even better than, the $R$-parameter in predicting solar eruptivity.

The linear correlation coefficient for the two sets of points in Fig.~\ref{boxplot} is 0.79 (SCC$=$0.53). If we had considered the BF gauge instead, that plot would have been only slightly different, and the respective correlation coefficient would be slightly larger, having the value 0.85 (SCC$=$0.59). In other words, the correlation of the two $R$ parameters is only weakly sensitive to the gauge dependency of RFLH. Moreover, if we consider separately the group of ARs with $R_\mathrm{H}<1.5\,10^5$~G and the one with $R_\mathrm{H}>1.5\,10^5$~G, we then find that the correlation coefficients between $R$ and $R_\mathrm{H}$ are 0.45 (SCC$=$0.41) and 0.99 (SCC$=$0.98), respectively, with both values referring to the DV gauge. This means that the correlation of the two $R$ parameters is stronger for ARs with higher values of $R_\mathrm{H}$, that is, of relative helicity contained in the MPIL, $H_\mathrm{MPIL}$.

\section{Discussion}
\label{sect:discussion}

This work dealt with the question whether relative field line helicity can be used as an indicator for solar eruptivity. To answer this question, we computed the NLFF extrapolated magnetic fields in a large number of snapshots from seven ARs of Solar Cycle 24, and from these we computed RFLH. We focused our study on the behaviour of RFLH in intervals of $\pm$~1h around 22 strong flares, above the M1.0 class, and in regions around the polarity inversion lines of the magnetic field (MPIL), and of RFLH as well (HPIL$_\mathrm{DV}$ and HPIL$_\mathrm{BF}$). In these PILs a Gaussian mask was employed in order to construct the regions of interest.

The use of RFLH allowed us to compute the relative helicity contained in the various PILs, something not possible otherwise. We found that in most of the flares the relative helicity of the MPILs (red curves in Figs.~\ref{dataplot2}-\ref{speplot2}) exhibits qualitatively similar temporal profiles as the relative helicity computed from the most accurate volume method. Moreover, when their absolute values are epochally superposed, a clear decrease of the MPIL helicities is observed, which is stronger than the respective decrease of the volume helicities. The MPIL helicities can also be used to define an equivalent parameter to the $R$ parameter of \citet{schrijver07}, simply by taking their square root and then dividing by the square of the pixel size. The values of this helicity-based parameter, $R_\mathrm{H}$, were found to correlate well with those of the original $R$-parameter, especially for the highest $R_\mathrm{H}$ values. Since high values of $R$ are related with strong flares, the same applies to $R_\mathrm{H}$ as well. In fact, the latter exhibit in some cases even higher values than the original parameter. This can be explained from the fact that helicity is more indicative of eruptivity than the magnetic flux, as has been also shown in, e.g., \citet{liokati22} or \citet{thalmann22}. Additionally, the $R_\mathrm{H}$ parameters resulting from the MPIL helicities experience larger absolute decreases during flares compared to the original $R$ parameter, adding to their value as possible markers of eruptivity.

The quality of the MPIL helicities as eruptivity indicators has to do with their association with the core structure involved in the eruption, and also, with their independence from the noisy helicity values at the vicinity of the AR. The larger changes during flares experienced by the MPIL helicities may be attributed to the reconfiguration of the magnetic field that takes place there, during reconnection, which removes helicity, and this is stronger around the MPIL. Moreover, it seems that the MPIL helicities, and the respective $R_\mathrm{H}$ parameters, may be used as early precursors of flares, since their decreases start 20-30~min before the flares. Hints for this can also be found in some models, \citep[e.g.,][]{pariat23}, where relative helicity decreases are observed earlier than the eruptive events.

The relative helicity contained in the HPILs (blue curves in Figs.~\ref{dataplot2}-\ref{speplot2}) is less indicative of upcoming eruptivity compared to the MPIL ones, although in many cases important flare-related changes are observed. This is evident from the incoherence of the temporal profiles of the HPIL helicity around flares, as in some cases they follow the respective profiles of the volume helicities and in others they do not. Additionally, when epochally superposed, the HPIL helicities show strong increases during the flares (in absolute values), unlike the volume or the MPIL helicities which decrease during flares. This incoherence is also depicted in the respective superposed $R_\mathrm{H}$ profiles of the HPIL helicities, which show little to no change during flares. It seems therefore that the HPIL relative helicities are not linked with solar eruptivity.

The limitation of the HPIL helicities stems from the fact that at the locations where RFLH changes sign the magnetic field has no special properties in principle, contrary to what happens at the MPIL. In other words, the HPILs are not that physically important as the respective magnetic PILs. The HPIL helicities therefore have no systematic evolution at the time of, and around, solar flares, and this leads to the observed lack of correlation with eruptions.

An issue that should be taken into account when examining RFLH is that, in principle, it is a gauge-dependent quantity. We addressed this in our work by examining two independent gauge setups for the vector potentials. The first involved the \citet{devore00} gauge, and the other the original \citet{BergerF84} one. Although the exact RFLH morphology shows a mild dependence on the gauge used, our results about $R_\mathrm{H}$ are not that sensitive to it. An estimate of the effect of the gauge can be derived from Fig.~\ref{speplot2} if we take the difference of the DV and BF curves and normalize it to the corresponding average of the two curves. For the case of the MPIL $R_\mathrm{H}$'s, this results in a $\sim 8\%$ difference, while for the HPIL $R_\mathrm{H}$'s this is much higher, $\sim 26\%$.

A factor not examined in our analysis is whether our results depend on a flare being eruptive or not. The reason is that our sample is not representative of this, since most confined flares in the sample come from a single AR, namely AR 12192. This issue could be the subject of a future study with a wider, and more balanced AR sample. We note however that preliminary analysis of the flares in our sample points to the same direction as the work of \citet{liu23}, in that the helicity changes during eruptive flares are more pronounced compared to those of the non-eruptive ones. Although the agreement of the two studies is reasonable given that the X-class flares in our sample is a subset of that in \citet{liu23}, the same difference in behaviour between eruptive and confined flares is also seen in \citet{duan23} and \citet{wang23}, with different AR samples. In the latter work, the authors find $\sim -15\%$ average changes of relative helicity during eruptive flares, and only $\sim -1\%$ for confined flares. Despite differences in methodology, all these studies point to the importance of helicity changes during solar flares, with our work highlighting the helicity changes at the location of the MPIL.

Another topology-related quantity that has recently demonstrated potential in predicting solar flares is magnetic winding \citep{raphaldini22}. This is a different measure of helicity where the flux weighting has been removed. In that work the magnetic winding performed better than helicity, indicating that the topological complexity of the magnetic field is more important than its flux. Our results seem to contradict \citet{raphaldini22}, as all PIL helicities are flux-weighted quantities. This discrepancy could be attributted to the different timescales involved in the two studies; in our case this is 2h, while in \citet{raphaldini22} it is a few days. Therefore, the use of helicity could be more appropriate when the timescale is shorter. In any case, quantities that have to do with the complexity of the magnetic field, weighted by flux or not, play the most important role when trying to predict solar eruptivity.

The focus of this work was on magnetic helicities, but many other parameters were computed apart from the $R_\mathrm{H}$'s, either related to the various PILs or to other regions of interest, such as the magnetic flux partitions produced by the method of \citet{georgoulis07}. The computed parameters involved not only magnetic helicity, but other physical quantities, such as the normal or tangential components of the electrical current on the photospheric plane. Their detailed study is left for the future, as the purpose of this paper was to show that there can be defined eruptivity indicators based on RFLH.

Another future direction of our work could include the examination of the behaviour of the current-carrying FLH, $h_\mathrm{j}$, defined in Eq.~(12) of \citet{moraitis19}. That term of RFLH could provide more insight into the flaring and/or eruptive processes of solar ARs, as the ratio of the respective component of helicity to the total relative helicity is known as an excelent eruptivity proxy, both in MHD simulations \citep{pariat17}, and in solar ARs \citep{moraitis19b}. 

Besides the direct forecasting of solar eruptivity, calculation of the changes of helicity over erupting PILs could help in supplying more accurate estimates of the magnetic helicity spawn by the associated CMEs, which could be used in assessments of the near-Sun CME magnetic field \citep[e.g.,][]{patsourakos16}, as well as in connecting with the magnetic helicity of the associated interplanetary CMEs \citep[e.g.,][]{mandrini04}. Linking the near-Sun conditions with those at the immediate environment of Earth is a difficult, but very important, problem, for which magnetic helicity could be a key quantity.

\begin{acknowledgements}
The authors thank the referee for carefully reading the paper and providing constructive comments. KM has received funding from the European Union’s Horizon 2020 research and innovation programme under the Marie Sk\l{}odowska-Curie grant agreement N$^o$ 893489. KM, SP, and AN also benefited from the ERC Whole Sun Synergy grant N$^o$ 810218. JT acknowledges the Austrian Science Fund (FWF): P31413-N27. EP acknowledges financial support by the French Programme National PNST of CNRS/INSU co-funded by CNES and CEA. The authors thank T. Wiegelmann for sharing his NLFF extrapolation code. They also acknowledge GRNET S.A. for awarding them access to the Greek national high performance computing system, ARIS.
\end{acknowledgements}

\bibliographystyle{aa}
\bibliography{refs}

\begin{appendix}

\section{Quality of the produced magnetic fields}
\label{app1}

To quantify the quality of the produced NLFF extrapolated magnetic fields we employ three different sets of metrics. First, metrics that quantify the solenoidality of the magnetic fields. We use three quantities for that purpose; the average absolute fractional flux increase, $\langle|f_i|\rangle$, defined in \citet{wheatland00}, the non-dimensional flux imbalance ratio
\begin{equation}
\epsilon_\mathrm{flux}=\frac{|\Phi^+-\Phi^-|}{(\Phi^++\Phi^-)},
\end{equation}
where $\Phi^+$ ($\Phi^-$) is the incoming (outgoing) flux entering through (leaving from) all the boundaries, and finally, the energy ratio $E_\mathrm{div}/E$, defined in \citet{valori16}. Here, $E_\mathrm{div}$ stands for the energy associated with the last three, non-physical terms in the following decomposition of the total magnetic energy:
\begin{equation}
E=E_\mathrm{p,s}+E_\mathrm{j,s}+E_\mathrm{p,ns}+E_\mathrm{j,ns}+E_\mathrm{mix},
\label{eq:nrgy}
\end{equation}
which results from splitting the magnetic field into potential and current-carrying components, and further dividing each component into solenoidal and non-solenoidal parts \citep{val13}. In a perfectly solenoidal magnetic field, all non-strictly solenoidal metrics vanish, and the total energy reduces to the usual form $E=E_\mathrm{p,s}+E_\mathrm{j,s}$, with $E_\mathrm{p,s}$ and $E_\mathrm{j,s}$ the potential and free energies respectively. The closer to zero these metrics are, the more solenoidal the magnetic field is. In practice, we use the threshold of $E_\mathrm{div}/E\lesssim 0.05$ for a magnetic field to be considered solenoidal enough, following \citet{thalmann19a}, but other options are discussed in Appendix~\ref{app3}.

The second set consists of two metrics that quantify the force-freeness of the magnetic fields. The first is the average current-weighted angle between the current and magnetic field, $\theta_J$, defined in \citet{wheatland00}, and the second, the average of the Lorentz force relative to its components
at each point of the domain, $\xi$, defined in \citet{malanushenko14}. The closer to zero these values are, the more force-free the magnetic field is.

The third set of metrics quantifies how well the computed vector potentials reconstruct their respective magnetic fields. For this we use the two most sensitive of the metrics given in \citet{schrijver06}, namely, the complement of the normalized vector error, $E'_n=1-E_n$, and the complement of the mean vector error, $E'_m=1-E_m$. We apply these separately to the two pairs ($\nabla\times\mathbf{A}, \mathbf{B}$) and ($\nabla\times\mathbf{A}_\mathrm{p}, \mathbf{B}_\mathrm{p}$), and so end up with four values. The closer to 1 these are, the better the reconstructions are.

\begin{figure}[h]
\centering
\includegraphics[width=0.46\textwidth]{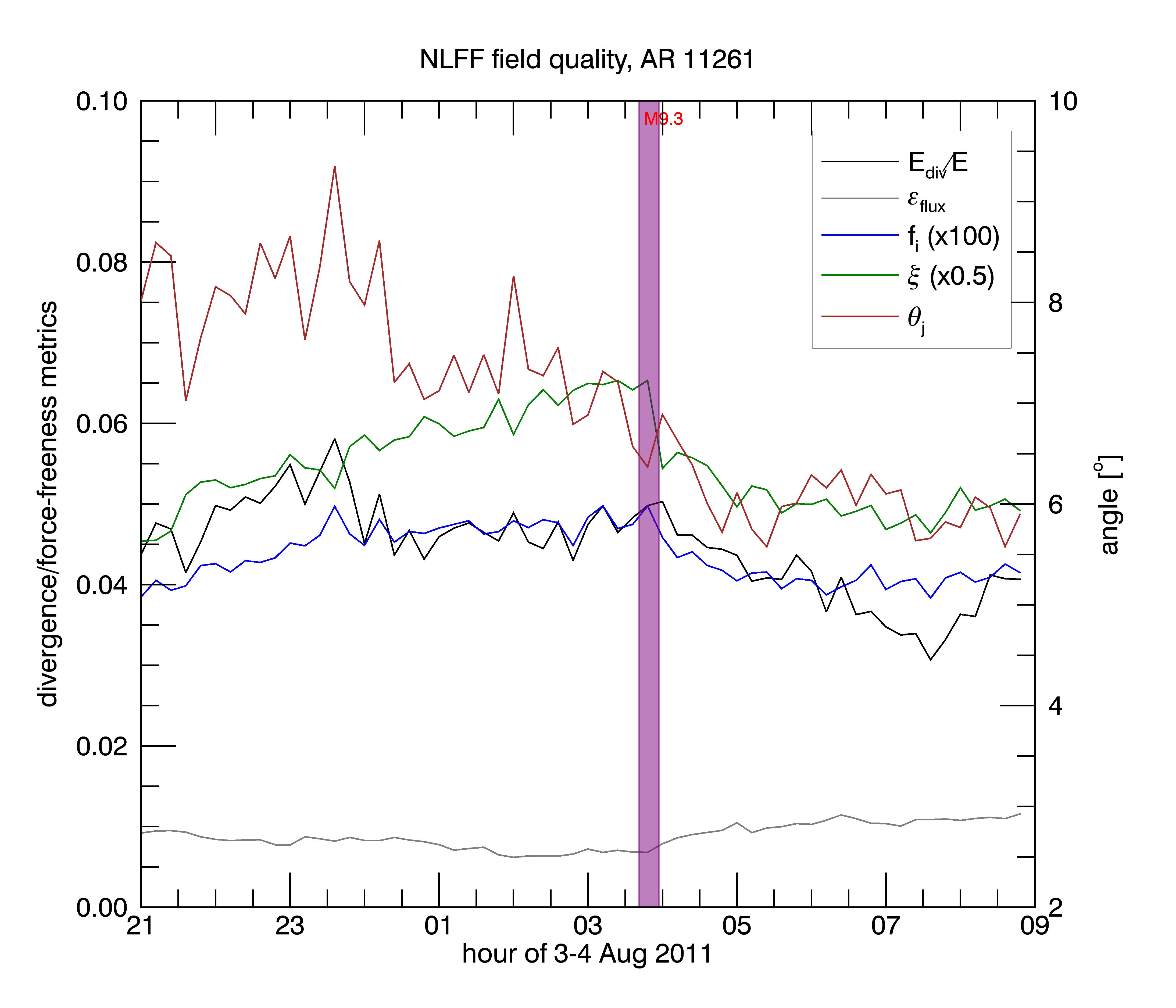}\\
\includegraphics[width=0.46\textwidth]{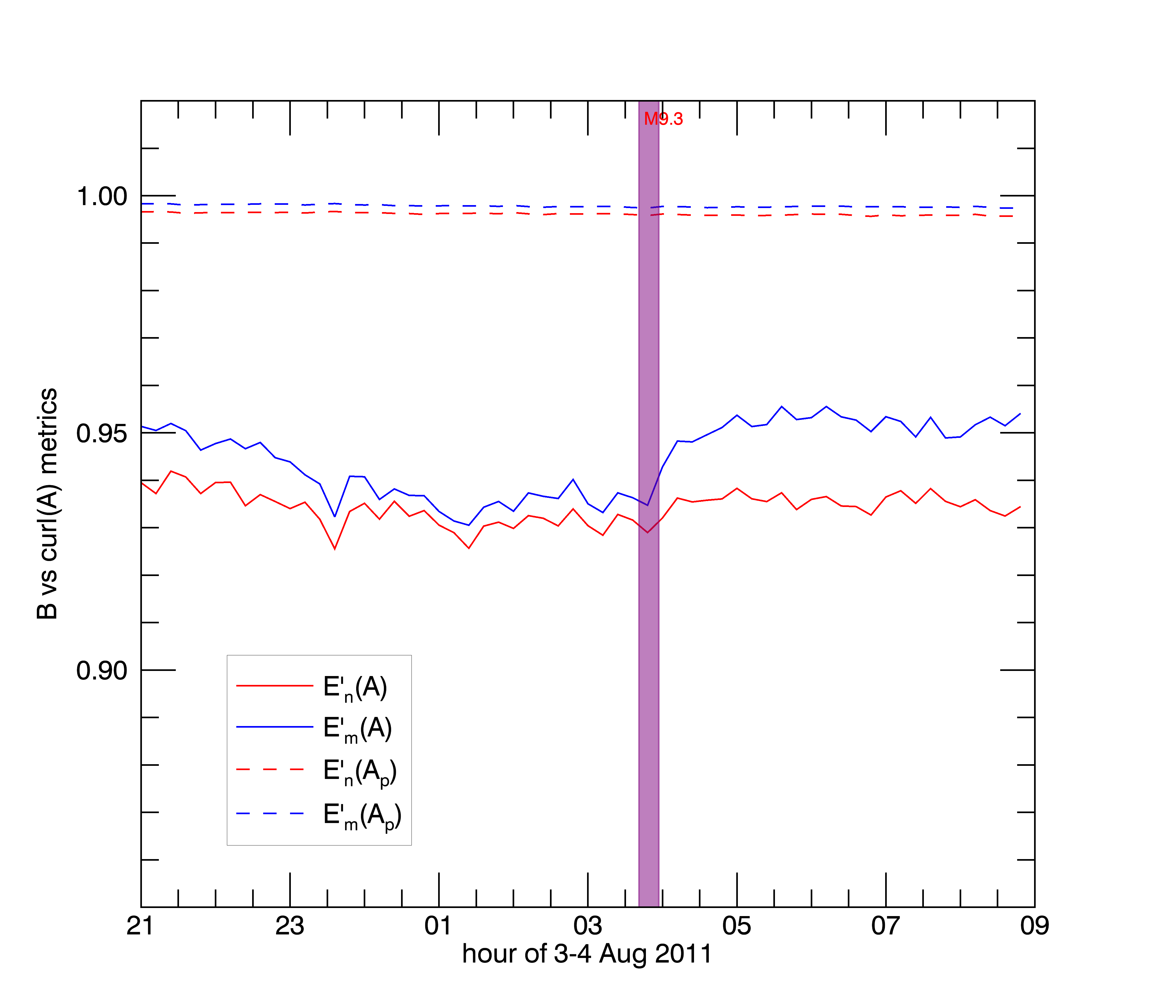}
\caption{Evolution of the divergence and force-freeness metrics (top), and of the $\mathbf{B}$ versus $\nabla\times\mathbf{A}$ metrics (bottom), for AR 11261. The purple bands denote the duration of the M9.3 flare of the AR.}
\label{metricsplot}
\end{figure}

An example of the evolution of these metrics for AR 11261 is shown in Fig.~\ref{metricsplot}. We note that all metrics indicate the high solenoidality and force-freeness of the produced NLFF fields. Focusing on $E_\mathrm{div}/E$ (black curve in Fig.~\ref{metricsplot}), we note that it is below the threshold of 0.05 at all times except from a few values around 23h of 3 August. The other ARs exhibit similar behaviour, with intervals where $E_\mathrm{div}/E$ is below the threshold and others where it is above it. As mentioned in the main text however, in Table~\ref{tab2}, at the times around the 22 flares examined in this work, $E_\mathrm{div}/E$ is well below 0.05. The time-averaged values of all metrics for the flares examined in this work are given in Table~\ref{tab4}. The interested reader can find the full list of the individual values of the metrics for all snapshots of the ARs at \url{doi.org/10.5281/zenodo.7607307}.

\begin{table*}
\caption{Time-averaged quality metrics for the NLFF fields of all examined flares and their corresponding standars errors.}
\centering
\resizebox{\textwidth}{!}{
\begin{tabular}{cccccccccc}
\hline
flare number & $\langle|f_i|\rangle\times10^4$ & $\epsilon_\mathrm{flux}\times10^3$ & ($E_\mathrm{div}/E)\times10^2$ & $\theta_J [^o]$ & $\xi$ & $E'_n(\mathbf{A})$ & $E'_m(\mathbf{A})$ & $E'_n(\mathbf{A}_\mathrm{p})$ & $E'_m(\mathbf{A}_\mathrm{p})$ \\
\hline
01 & 3.80$\pm$0.19 & 1.90$\pm$0.13 & 0.23$\pm$0.03 & 15.70$\pm$0.40 & 0.1800$\pm$0.0100 & 0.9650$\pm$0.0010 & 0.9550$\pm$0.0027 & 0.98770$\pm$0.00070 & 0.97700$\pm$0.00300 \\
02 & 2.40$\pm$0.03 & 1.70$\pm$0.13 & 0.50$\pm$0.03 & 13.53$\pm$0.09 & 0.0910$\pm$0.0019 & 0.9753$\pm$0.0001 & 0.9759$\pm$0.0004 & 0.98990$\pm$0.00060 & 0.98200$\pm$0.00290 \\
03 & 4.64$\pm$0.08 & 7.50$\pm$0.30 & 4.73$\pm$0.07 & 6.80$\pm$0.14 & 0.1220$\pm$0.0028 & 0.9324$\pm$0.0008 & 0.9400$\pm$0.0018 & 0.99606$\pm$0.00004 & 0.99763$\pm$0.00004 \\
04 & 4.22$\pm$0.08 & 7.10$\pm$0.20 & 3.80$\pm$0.19 & 8.30$\pm$0.15 & 0.0950$\pm$0.0014 & 0.8960$\pm$0.0011 & 0.9110$\pm$0.0010 & 0.99801$\pm$0.00002 & 0.99879$\pm$0.00001 \\
05 & 5.61$\pm$0.06 & 1.50$\pm$0.26 & 2.00$\pm$0.13 & 7.10$\pm$0.18 & 0.1220$\pm$0.0011 & 0.9168$\pm$0.0007 & 0.9312$\pm$0.0006 & 0.99300$\pm$0.00140 & 0.99000$\pm$0.00400 \\
06 & 4.75$\pm$0.06 & 1.80$\pm$0.23 & 0.86$\pm$0.08 & 5.80$\pm$0.11 & 0.0930$\pm$0.0011 & 0.9331$\pm$0.0007 & 0.9380$\pm$0.0010 & 0.98990$\pm$0.00090 & 0.98100$\pm$0.00230 \\
07 & 4.47$\pm$0.02 & 3.50$\pm$0.16 & 0.53$\pm$0.07 & 7.50$\pm$0.15 & 0.1015$\pm$0.0006 & 0.9358$\pm$0.0006 & 0.9443$\pm$0.0005 & 0.98310$\pm$0.00060 & 0.96400$\pm$0.00140 \\
08 & 3.90$\pm$0.05 & 6.50$\pm$0.20 & 0.53$\pm$0.05 & 7.90$\pm$0.23 & 0.0915$\pm$0.0009 & 0.9410$\pm$0.0008 & 0.9528$\pm$0.0007 & 0.97150$\pm$0.00070 & 0.93500$\pm$0.00190 \\
09 & 2.48$\pm$0.03 & 3.72$\pm$0.07 & 4.70$\pm$0.13 & 4.97$\pm$0.08 & 0.0692$\pm$0.0008 & 0.9447$\pm$0.0006 & 0.9575$\pm$0.0004 & 0.99799$\pm$0.00004 & 0.99881$\pm$0.00002 \\
10 & 2.79$\pm$0.03 & 4.00$\pm$0.15 & 3.97$\pm$0.08 & 5.80$\pm$0.11 & 0.0699$\pm$0.0005 & 0.9407$\pm$0.0006 & 0.9425$\pm$0.0009 & 0.99826$\pm$0.00002 & 0.99910$\pm$0.00001 \\
11 & 2.98$\pm$0.04 & 3.05$\pm$0.20 & 3.60$\pm$0.14 & 6.40$\pm$0.20 & 0.0800$\pm$0.0004 & 0.9375$\pm$0.0009 & 0.9354$\pm$0.0009 & 0.99831$\pm$0.00003 & 0.99916$\pm$0.00001 \\
12 & 3.66$\pm$0.03 & 0.51$\pm$0.03 & 1.98$\pm$0.08 & 5.64$\pm$0.03 & 0.1060$\pm$0.0005 & 0.9473$\pm$0.0003 & 0.9487$\pm$0.0003 & 0.99578$\pm$0.00009 & 0.99570$\pm$0.00019 \\
13 & 3.60$\pm$0.02 & 0.11$\pm$0.03 & 2.11$\pm$0.06 & 5.65$\pm$0.05 & 0.1080$\pm$0.0007 & 0.9489$\pm$0.0003 & 0.9510$\pm$0.0003 & 0.99677$\pm$0.00006 & 0.99790$\pm$0.00012 \\
14 & 3.60$\pm$0.01 & 0.11$\pm$0.02 & 1.88$\pm$0.05 & 5.52$\pm$0.03 & 0.1093$\pm$0.0004 & 0.9490$\pm$0.0002 & 0.9516$\pm$0.0003 & 0.99677$\pm$0.00005 & 0.99790$\pm$0.00013 \\
15 & 3.49$\pm$0.02 & 0.13$\pm$0.02 & 2.22$\pm$0.03 & 5.59$\pm$0.05 & 0.1092$\pm$0.0003 & 0.9517$\pm$0.0003 & 0.9536$\pm$0.0004 & 0.99688$\pm$0.00005 & 0.99800$\pm$0.00012 \\
16 & 3.62$\pm$0.01 & 1.08$\pm$0.02 & 2.07$\pm$0.03 & 5.70$\pm$0.03 & 0.1050$\pm$0.0002 & 0.9502$\pm$0.0002 & 0.9449$\pm$0.0002 & 0.99494$\pm$0.00007 & 0.99360$\pm$0.00016 \\
17 & 3.68$\pm$0.02 & 1.27$\pm$0.04 & 1.79$\pm$0.06 & 5.56$\pm$0.04 & 0.1065$\pm$0.0003 & 0.9484$\pm$0.0003 & 0.9436$\pm$0.0003 & 0.99450$\pm$0.00013 & 0.99260$\pm$0.00028 \\
18 & 3.71$\pm$0.03 & 0.89$\pm$0.05 & 2.13$\pm$0.05 & 5.65$\pm$0.03 & 0.1089$\pm$0.0005 & 0.9465$\pm$0.0003 & 0.9393$\pm$0.0004 & 0.99539$\pm$0.00008 & 0.99450$\pm$0.00019 \\
19 & 3.94$\pm$0.01 & 0.12$\pm$0.02 & 1.83$\pm$0.02 & 5.39$\pm$0.03 & 0.1128$\pm$0.0004 & 0.9421$\pm$0.0001 & 0.9310$\pm$0.0003 & 0.99664$\pm$0.00006 & 0.99710$\pm$0.00013 \\
20 & 4.02$\pm$0.03 & 0.06$\pm$0.01 & 0.79$\pm$0.04 & 6.19$\pm$0.03 & 0.1118$\pm$0.0006 & 0.9409$\pm$0.0003 & 0.9326$\pm$0.0005 & 0.99625$\pm$0.00004 & 0.99630$\pm$0.00010 \\
21 & 4.48$\pm$0.04 & 0.35$\pm$0.01 & 1.15$\pm$0.08 & 5.56$\pm$0.02 & 0.1117$\pm$0.0003 & 0.9287$\pm$0.0009 & 0.9204$\pm$0.0009 & 0.99699$\pm$0.00003 & 0.99785$\pm$0.00007 \\
22 & 4.73$\pm$0.06 & 3.30$\pm$0.14 & 4.00$\pm$0.50 & 8.50$\pm$0.80 & 0.0997$\pm$0.0009 & 0.9270$\pm$0.0013 & 0.9220$\pm$0.0026 & 0.99720$\pm$0.00010 & 0.99804$\pm$0.00007 \\
\hline
\end{tabular}
}
\label{tab4}
\end{table*}

\section{Dependence of the results on the value of AR pixel size}
\label{app2}

An issue of concern might be the different pixel sizes between ARs, since all ARs have a pixel size of 1", except from AR 11158 and AR 12192, where it is 2". To check whether this difference produces an important effect, we performed NLFF extrapolations (with the same parameters as in Sect.~\ref{sect:method}) on a subset of AR's 11158 magnetograms that were rebinned to the pixel size of 1". We chose the period during the X2.2 flare of 15 February 2011, and used 12 snapshots with HMI's highest cadence of 12 minute, from 00:48 UT to 03:00 UT. 

\begin{figure}[h]
\centering
\includegraphics[width=0.46\textwidth]{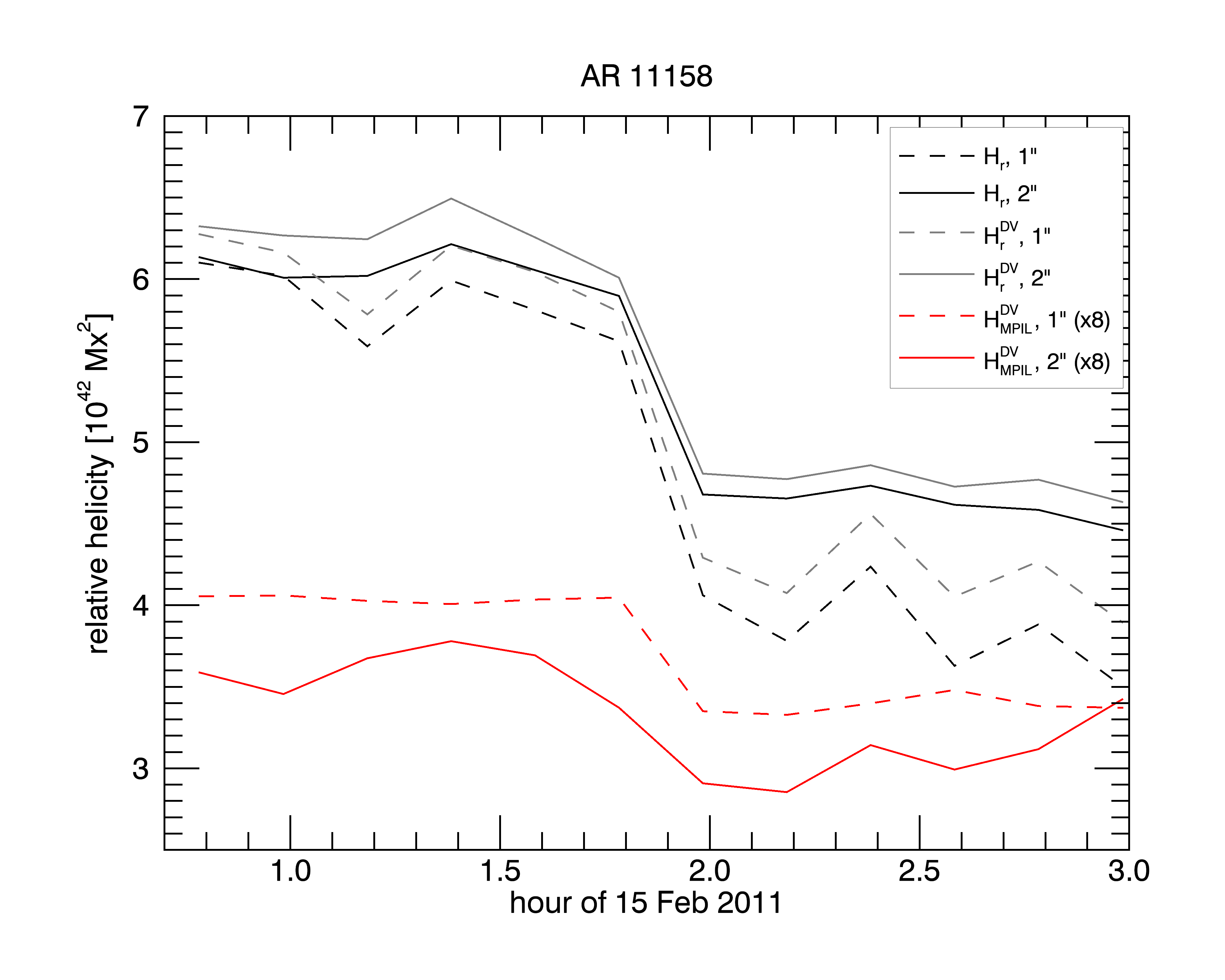}
\caption{Evolution of three relative helicities, $H_\mathrm{r}$, $H_\mathrm{r}^\mathrm{DV}$, and $H_\mathrm{MPIL}^\mathrm{DV}$, during the X2.2 flare of AR 11158, when the NLFF fields have a pixel size of 1" (dashed curves), and 2" (solid curves).}
\label{app2plot}
\end{figure}

We computed various relative helicities for the 1" NLFF fields and show in Fig.~\ref{app2plot} the comparison of some of them with the respective 2" results. We note that the different pixel size produces helicities of similar trend but of different magnitude. The average difference between the 1" and 2" helicities in Fig.~\ref{app2plot} normalized to their mean value is $\sim 10\%$. We thus expect that the overall effect in our results of the different pixel size may be estimated to the same amount, $\sim 10\%$.

\section{Dependence of the results on the value of the threshold for $E_\mathrm{div}/E$}
\label{app3}

The value of the parameter $E_\mathrm{div}/E$ is critical for the quality of relative helicity computations. This has been first shown in \citet{valori16} where a threshold of $E_\mathrm{div}/E<0.08$ was suggested for reliable helicity values. Later, \citet{thalmann19a} proposed an even lower threshold of 0.05. The latter value is used in the main part of this work, and here we examine the effect that the different values of $E_\mathrm{div}/E$ have on our results. 

The original sample of 22 flares for which $E_\mathrm{div}/E<0.05$ is denoted as S2 in this Appendix. When we increase the threshold to $E_\mathrm{div}/E<0.085$, the number of flares is increased by four eruptive flares of AR 11890, namely an X3.3 with peak time at 22:12 UT of 5 November 2013 and average $E_\mathrm{div}/E=(5.70\pm 0.17)\times10^{-2}$ during the flare, an M3.8 (13:46 UT of 6 November 2013, $E_\mathrm{div}/E=(4.9\pm 0.2)\times10^{-2}$), an X1.1 (04:26 UT of 8 November 2013, $E_\mathrm{div}/E=(5.03\pm 0.05)\times10^{-2}$), and another X1.1 (05:14 UT of 10 November 2013, $E_\mathrm{div}/E=(8.10\pm 0.15)\times10^{-2}$). This sample is denoted as S1. When the threshold is decreased to $E_\mathrm{div}/E<0.025$, the number of flares reduces to 16, since flares with numbers 3-4, 9-11, and 22 in Table~\ref{tab2} drop out. This sample consists of only three ARs, 11158, 11618, and 12192, and is denoted as S3. The characteristics of the three samples are summarized in Table~\ref{tab5}.

\begin{table}[ht]
\caption{Characteristics of the three ARs samples considered.}
\centering
\resizebox{0.48\textwidth}{!}{
\begin{tabular}{cccccc}
\hline
sample & threshold & ARs & flares & X flares & eruptive flares \\
\hline
S1 & 0.085 & 7 & 26 & 8 & 15 \\
S2 & 0.050 & 7 & 22 & 5 & 11 \\
S3 & 0.025 & 3 & 16 & 3 & 7 \\
\hline
\end{tabular}
}
\label{tab5}
\end{table}

\begin{figure}[h]
\centering
\includegraphics[width=0.46\textwidth]{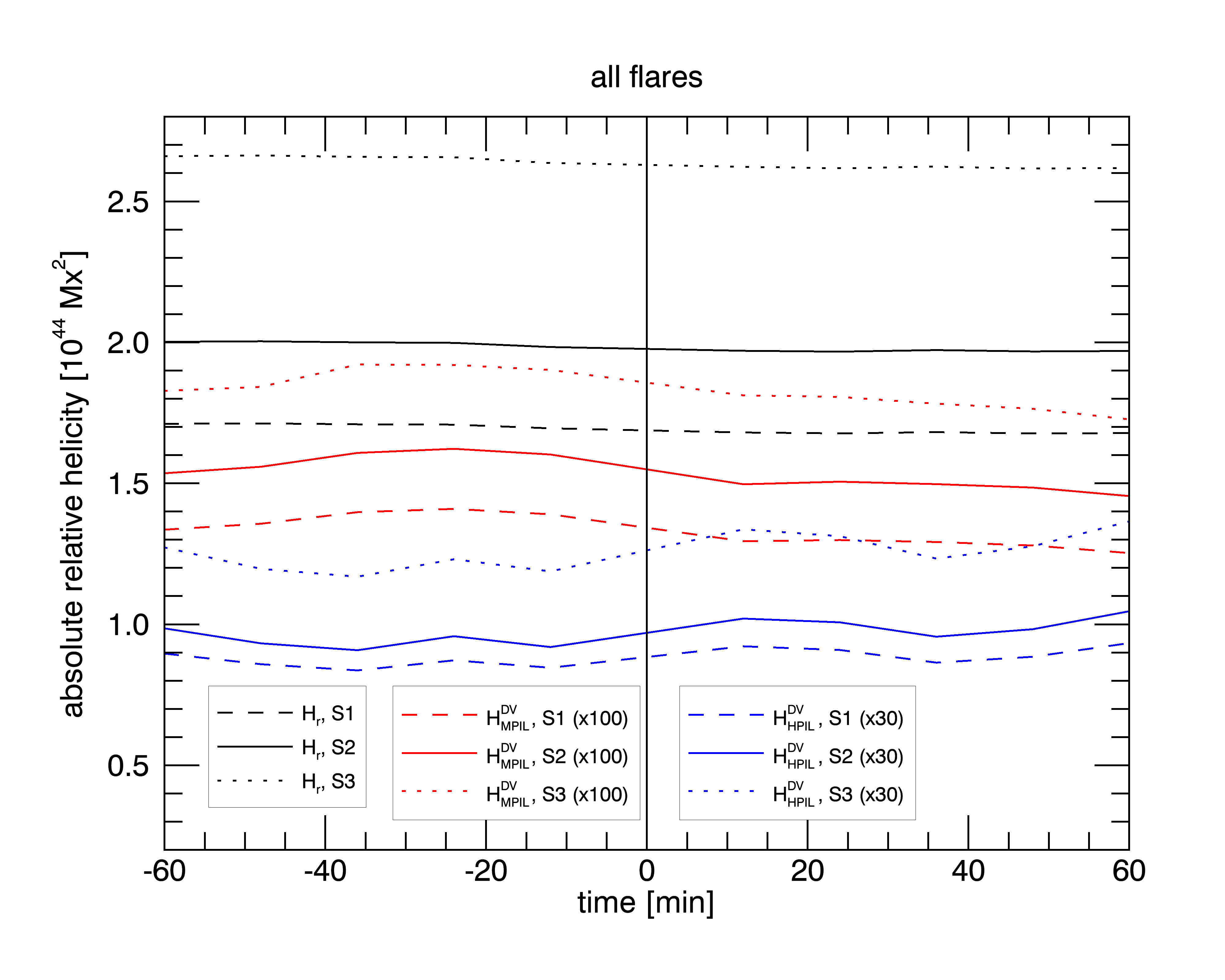}
\caption{Relative helicity SPE profiles for $H_\mathrm{r}$, $H_\mathrm{MPIL}^\mathrm{DV}$, and $H_\mathrm{HPIL}^\mathrm{DV}$ as in Fig.~\ref{speplot}, for the three examined ARs samples.}
\label{app3plot}
\end{figure}

To examine the effect of varying $E_\mathrm{div}/E$, we focus on three of the curves of Fig.~\ref{speplot}, namely on $H_\mathrm{r}$, $H_\mathrm{MPIL}^\mathrm{DV}$, and $H_\mathrm{HPIL}^\mathrm{DV}$. We repeat the computations described in Sect.~\ref{sect:res2} for S1 and S3, and show the results for these helicities in the three samples in Fig.~\ref{app3plot}. We notice that the qualitative behaviour of all helicity profiles is not affected by the value of $E_\mathrm{div}/E$, only the relative normalization changes due to the different sample sizes. The SPE profile for $H_\mathrm{r}$ shows a mild decrease during flares which becomes even milder as $E_\mathrm{div}/E$ decreases, that is, as the sample size decreases. More specifically, the average decreases drop from -1.7\% for S1 to -1.3\% for S3. The MPIL helicities show the strongest decreases, which become less intense with decreasing $E_\mathrm{div}/E$ (on average -6.8\% for S1 and -5.5\% for S3). The HPIL helicity profiles on the other hand increase during the flares, with larger magnitude as $E_\mathrm{div}/E$ decreases (on average 4.7\% for S1 and 7.7\% for S3). Moreover, all these behaviours hold for both the instantaneous, and the average changes.

\end{appendix}

\end{document}